\documentclass[aps,pra,amsmath,superscriptaddress,floatfix,notitlepage,balancelastpage,twocolumn,reprint]{revtex4}
\usepackage[english]{babel}
\usepackage[utf8]{inputenc}
\usepackage{amsmath}
\usepackage{graphicx}
\usepackage[colorinlistoftodos]{todonotes}
\usepackage{bibentry}
\usepackage{url}
\usepackage{rotating}
\usepackage{float}
\usepackage{color}
\usepackage[sort&compress]{natbib}
\usepackage{hyperref}
\usepackage[caption=false]{subfig}
\usepackage{empheq}
\usepackage{lipsum}
\hypersetup{colorlinks,citecolor={red}}
\newcommand{\grad}{\boldsymbol{\nabla}}
\newcommand{\br}{\boldsymbol{r}}
\newcommand{\bv}{\boldsymbol{v}}
\newcommand{\rhoh}{\hat{\rho}}
\newcommand{\curO}{{\cal O}}
\newcommand{\curA}{{\cal A}}
\newcommand{\curV}{{\cal V}}
\newcommand{\curN}{{\cal N}}
\newcommand{\curS}{{\cal S}}
\newcommand{\be}{\begin{equation}}
\newcommand{\ee}{\end{equation}}
\newcommand{\bea}{\begin{eqnarray}}
\newcommand{\eea}{\end{eqnarray}}

\begin{document}
\title{Inflationary Dynamics and Particle Production
  in a Toroidal Bose-Einstein Condensate}

\author{Anshuman Bhardwaj}
\email{abhard4@lsu.edu}
\author{Dzmitry Vaido}
\email{dvaido1@lsu.edu}
\author{Daniel E. Sheehy}
\email{sheehy@lsu.edu}
\affiliation{Department of Physics and Astronomy, Louisiana State University, Baton Rouge, LA 70803 USA}

\date{Jan 29, 2021}

\begin{abstract}
  We present a theoretical study of the dynamics of a Bose-Einstein
  condensate (BEC) trapped inside an expanding toroid that can realize
  an analogue inflationary universe.   As the system expands, we find
  that phonons in the BEC undergo redshift and damping due to quantum
  pressure effects, owing to the thinness of the ring.  We predict that rapidly expanding toroidal BEC's
  can exhibit spontaneous particle creation, and study this phenomenon
  in the context of an initial coherent state wavefunction.  We show
  how particle creation would be revealed in the atom density and
  density correlations, and discuss connections to the cosmological
  theory of inflation. 
\end{abstract}

\maketitle 

\section{Introduction} The theory of inflation is the most promising description of the early
universe~\textcolor{red}{\cite{Starobinsky:1980te,Guth:1980zm,Albrecht:1982wi,Hawking:1981fz,Linde:1981mu,Linde:1983gd,Liddle:2000cg}},
although alternatives exist~\textcolor{red}{\cite{Ijjas:2018qbo}}.
This theory is based on a 
field $\phi$, the inflaton, propagating in a classical
spacetime and moving under the influence of its own potential
$V(\phi)$.
The quantum fluctuations in the inflaton field couple with the spatial
curvature of the universe, thus acting as seeds for the observed
cosmic microwave background (CMB)
anisotropies~\textcolor{red}{\cite{Ade:2015xua}} and the large scale
structure of our universe~\textcolor{red}{\cite{Tegmark:2003ud}},
although primordial gravitational waves are yet to be observed.
The exact shape of the  potential $V(\phi)$ is currently not known,
although work has been done to reconstruct
it~\textcolor{red}{\cite{Lidsey:1995np}}. Indeed, experiments have
put stringent constraints on some of the candidates such as the
quadratic and quartic inflationary potentials, though there remains a huge class
of models that are able to explain 
observations~\textcolor{red}{\cite{Ade:2015lrj}}.  In addition, the CMB
observations  have revealed possible anomalies on the largest
scales with a $3\sigma$ significance, that hint towards 
new physics~\textcolor{red}{\cite{Ade:2015hxq}}. However, testing
inflationary models using cosmological experiments is expensive and
difficult.

A natural question, then, is if there exists an alternative setting to
test the predictions of inflation.
The answer is `Analogue
Gravity'~\textcolor{red}{\cite{Barcelo:2005fc}}, where the aim is to
come up with simple experimental setups that can be performed in a lab
and which mimic the equations governing gravitational and cosmological 
phenomena such as inflation and black hole physics.  Early work in 
this direction came from Unruh who, in 1981, 
showed~\textcolor{red}{\cite{Unruh:1980cg}} that the Navier-Stokes'
equations for fluid flow, such as in a draining bathtub, could mimic Hawking
radiation~\textcolor{red}{\cite{Hawking:1974rv,Hawking:1974sw}}
coming from a black hole horizon.  This showed that analogue black holes can be
constructed, allowing the study of near-horizon physics outside of an astrophysical
setting. 
Several recent experiments 
have confirmed the existence of such  analogue Hawking
radiation~\textcolor{red}{\cite{Philbin:2007ji,Belgiorno:2010wn,Weinfurtner:2010nu,Steinhauer:2015saa}},
as well as other phenomena such as 
classical superradiance~\textcolor{red}{\cite{Torres:2016iee}}, the
Casimir effect~\textcolor{red}{\cite{Jaskula:2012ab}}, and
Sakharov oscillations~\textcolor{red}{\cite{Hung:2012nc}}.
Other analogue gravity proposals test ideas like the Gibbons-Hawking
effect~\textcolor{red}{\cite{Fedichev:2003id}}, the vacuum
decay~\textcolor{red}{\cite{Fialko:2014xba,Fialko:2016ggg,Braden:2017add,Billam:2018pvp}}  and the
Unruh effect~\textcolor{red}{\cite{Rodriguez-Laguna:2016kri}}.

\begin{figure}[ht!]
   \begin{center}
     \includegraphics[width=0.4\textwidth]{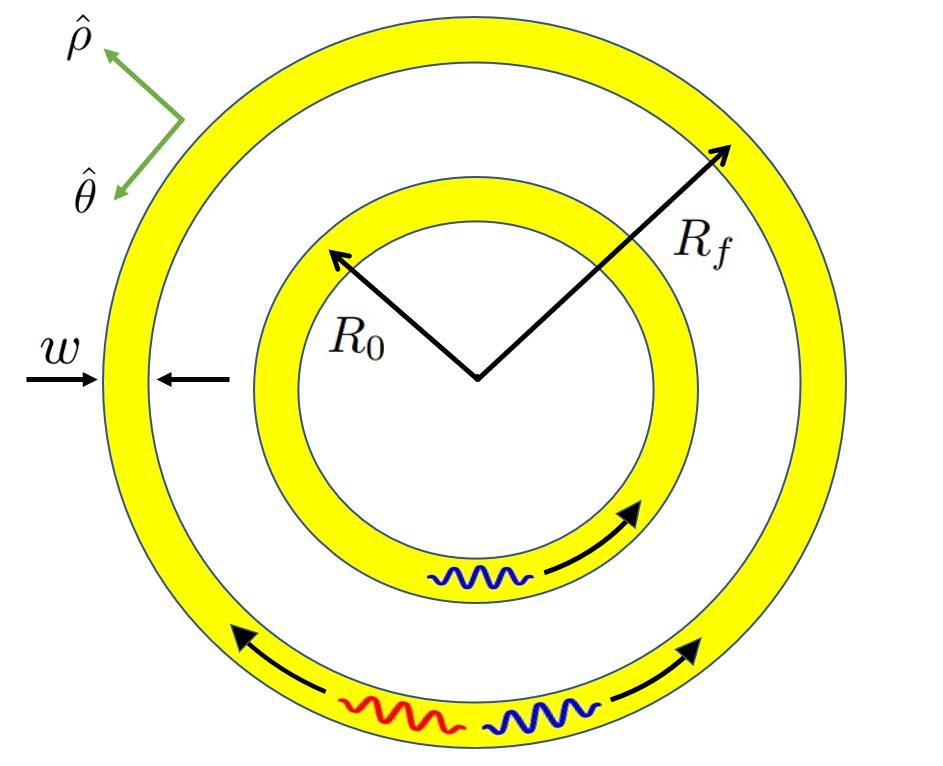}
     \end{center}
   \caption{(Color Online) Sketch of the initial (with radius $R_0$) and final
   (with radius $R_f$) state of an expanding ring-shaped (toroidal) BEC, as
     realized in Ref.~\onlinecite{Eckel:2017uqx}.  As depicted, an initial
     density wave traveling counterclockwise ({\em blue}) bifurcates due to phonon
     creation, into two  counter propagating waves ({\em red and blue}).
      In the main text, we use polar
      coordinates ($\rho,\theta,z$) with 
       $z$ directed out of the page.  
      We also indicated the ring diameter $w$, implying that the ring
     cross-sectional area $\curA \simeq
     \frac{1}{4}\pi w^2$ for the case of a circular cross section.
However,  in the main text we shall allow for different
radii in the $\rho$ and $z$ direction (denoted by $R_\rho$ and $R_z$, respectively).
   }
\label{expandingSketch}
\end{figure}

Inflation in analogue systems has been studied
theoretically~\textcolor{red}{\cite{Barcelo:2003wu,Fedichev:2003bv,Fischer:2004bf,Jain:2007gg,Prain:2010zq,Llorente:2019rbs,Fifer:2018hcv}},
and realized experimentally in Bose-Einstein
condensates (BECs)~\textcolor{red}{\cite{Eckel:2017uqx}} and in ion
traps~\textcolor{red}{\cite{Wittemer:2019kds}}.  Here, our primary motivation is the
BEC implementation of analogue inflation realized by  Eckel et al~\textcolor{red}{\cite{Eckel:2017uqx}}.
The Eckel et al experiments featured
a BEC in a time-dependent trap with the shape of a thin toroid,
with a rapid expansion of the toroid mimicking the inflationary era.
Some of the analogues of cosmological phenomena observed by Eckel et al
were the red-shifting of frequencies and damping of modes due to expansion.
Vortex creation after halting of the ring expansion was also observed,
an analogue of reheating in the early universe.

In this paper, we present analytical results to help understand the Eckel et al 
experiments and discuss other possible observables that can probe inflationary physics in
the context of an analogue BEC experiment. 
  In particular, we show that the thinness of the toroid introduces significant quantum pressure corrections in the BEC, that cause damping of sound modes and eventually lead to spontaneous phonon creation.
The rest of the paper is organized as follows:
in Sec.~\ref{sec:BDG} , we investigate the evolution of perturbations (i.e., phonons) in a BEC
using the Bogoliubov-de Gennes Hamiltonian.
In Sec.~\ref{sec:MSE}, we show that
this approach leads to the Mukhanov-Sasaki
equation~\textcolor{red}{\cite{Sasaki:1983kd,Kodama:1985bj,Mukhanov:1988jd}}
that governs the evolution of such phonons in the primordial
universe. This equation forces the density fluctuations  to undergo damping
that comes from quantum pressure.
In Sec.~\ref{sec:SPC}, we show that as the
ring undergoes an expansion there is a spontaneous generation of phonons
which is the analogue of particle creation in the early universe.
In Sec.~\ref{sec:STIM}, we study the case of stimulated phonon creation by calculating
the average density in a coherent state and
showing that an initial traveling density wave bifurcates (due to phonon
creation) into two waves, as illustrated in Fig.~\ref{expandingSketch}.
In Sec.~\ref{sec:DensityCorrelations}, we calculate the density-density
correlation function for this system at zero and finite temperatures, showing that the angle-dependence of
density correlations exhibits a signature of phonon production.
In Sec.~\ref{sec:concl}, we provide brief concluding remarks and in Appendix~\ref{appendix}
we provide details that are omitted from the main text.

\section{Bogoliubov-de Gennes Hamiltonian}
\label{sec:BDG}
In this section we describe, within the Bogoliubov-de Gennes (BdG)
formalism, how a boson gas in a time-dependent trap can exhibit
emergent relativistic dynamics that mimic the phenomenon of
inflation. We start with the following Hamiltonian that describes a
Bose-Einstein condensate (BEC), given in terms of a complex scalar
field ($\hat{\Phi}(\boldsymbol{r})$), evolving inside a non-uniform and
time dependent toroidal potential $V(\boldsymbol{r},t)$:
\begin{eqnarray}\label{BdGHamiltonian}
\hat{H} & = & \hat{H}_{0} + \hat{H}_{1}, \nonumber \\
\hat{H}_{0} & = & \int d^{3}r~\hat{\Phi}^{\dagger}(\boldsymbol{r})
\bigg[-\frac{\hbar^{2}}{2M}\nabla^{2}+V(\boldsymbol{r})-\mu\bigg]\hat{\Phi}(\boldsymbol{r}), \nonumber \\
\hat{H}_{1} & = & \frac{U}{2}\int d^{3}r~\hat{\Phi}^{\dagger}(\boldsymbol{r})\hat{\Phi}^{\dagger}(\boldsymbol{r})
\hat{\Phi}(\boldsymbol{r})\hat{\Phi}(\boldsymbol{r}),
\end{eqnarray}
where 
$\mu$  is the chemical potential, $U=\frac{4\pi a_{s}\hbar^{2}}{M}$ is
the interaction parameter, $a_{s}$ is the scattering length, $\hbar$
is Planck's constant, and $M$ is the mass of the bosonic atoms.  
The time-dependent single-particle potential $V(\boldsymbol{r},t)$ describes a time-dependent
toroidal potential which, taking a cylindrical coordinate system $\boldsymbol{r} = (\rho,\theta,z)$,
we can take to be parabolic in the $\hat{z}$ direction and
a higher power law in the
radial ($\hat{\rho}$) direction:
\begin{equation}
  \label{eq:veeofr}
V(\boldsymbol{r},t)=\frac{1}{2}M\omega_{z}^{2}z^{2}+\lambda|\rho-R(t)|^{n},
\end{equation}
consistent with the experiments of Eckel et al~\cite{Eckel:2017uqx}, who realize a ``flat bottomed'' trap with the exponent $n\simeq 4$.
Here, $R(t)$ is the externally-controlled radius of the toroid that increases with
time. The condensate field operator obeys the commutation relation
\begin{equation}
\big[\hat{\Phi}(\boldsymbol{r}),\hat{\Phi}^{\dagger}(\boldsymbol{r'})\big]
= \delta^{(3)}(\boldsymbol{r}-\boldsymbol{r'}).
\end{equation}
Under the BdG
approximation, the condensate field operator in the Heisenberg picture
$\hat{\Phi}(\boldsymbol{r},t)$ can be written as the sum of a coherent
background $\Phi_{0}(\boldsymbol{r},t)$ and the perturbation operator
$\delta\hat{\phi}(\boldsymbol{r},t)$:
\begin{equation}
  \label{eq:phideltaphi}
  \hat{\Phi}=\Phi_{0}(1+\delta\hat{\phi}).
\end{equation}
Plugging Eq.~(\ref{eq:phideltaphi}) into the
Hamiltonian (\ref{BdGHamiltonian}) and using Heisenberg's equations of
motion, we get
\begin{eqnarray}\label{BdGEquation}
\hspace{-.4cm}  i\hbar\partial_{t}\delta\hat{\phi}\!\!\! &=&\!\!\!
  -\frac{\hbar^{2}}{2M}\nabla^{2}\delta\hat{\phi}
  -\frac{\hbar^{2}}{M}\frac{\boldsymbol{\nabla}\Phi_{0}}{\Phi_{0}}\!\cdot\!\boldsymbol{\nabla}\delta\hat{\phi}
  \!+\! Un_{0}\big[\delta\hat{\phi}^{\dagger}\!\! +\! \delta\hat{\phi}\big], 
\end{eqnarray}
for the $\delta\hat{\phi}(\boldsymbol{r},t)$ equation of motion~\textcolor{red}{\cite{Llorente:2019rbs}}.
Here, we defined the background density as
$n_{0}(\boldsymbol{r},t)\equiv|\Phi_{0}(\boldsymbol{r},t)|^{2}$.

Equation~(\ref{BdGEquation}) describes dynamics of the perturbation operator
$\delta\hat{\phi}$
in the presence of a time-dependent background $\Phi_0(\boldsymbol{r},t)$.  Below, we find it convenient to transform to
the Madelung representation in terms  of density $\hat{n}(\boldsymbol{r},t)$ and phase $\hat{\phi}(\boldsymbol{r},t)$
field operators via:
\begin{equation}\label{Madelung}
\hat{\Phi}(\boldsymbol{r},t)=\sqrt{\hat{n}(\boldsymbol{r},t)}e^{i\hat{\phi}(\boldsymbol{r},t)}.
\end{equation}
To proceed we use Eq.~(\ref{eq:phideltaphi})
on the left hand side of Eq.~(\ref{Madelung})
and we introduce linear perturbations
 for the phase $\hat{\phi}=\phi_{0}+\hat{\phi}_{1}$
and the density $\hat{n}=n_{0}+\hat{n}_{1}$ on the
right hand side of Eq.~(\ref{Madelung}).  Then, keeping first-order contributions, we
obtain an
expression for the condensate perturbation $\delta\hat{\phi}$ in terms
of the perturbations in density $\hat{n}_{1}(\boldsymbol{r},t)$ and phase
$\hat{\phi}_{1}(\boldsymbol{r},t)$:
\begin{eqnarray}\label{RelnDelPhiandn1Phi1}
\delta\hat{\phi}(\boldsymbol{r},t) & = & \frac{\hat{n}_{1}(\boldsymbol{r},t)}{2n_{0}(\boldsymbol{r},t)}
+i\hat{\phi}_{1}(\boldsymbol{r},t).
\end{eqnarray}
Here $\phi_{0}(\boldsymbol{r},t)$ denotes the 
background phase.  The phase and density perturbations
satisfy the commutation relation
$\big[\hat{n}_{1}(\boldsymbol{r},t),\hat{\phi}_{1}(\boldsymbol{r'},t)\big]
= i\delta^{(3)}(\boldsymbol{r}-\boldsymbol{r'})$. Substituting the
above relation into (\ref{BdGEquation}) and using Madelung's
representation for the background
$\Phi_{0}(\boldsymbol{r},t)=\sqrt{n_{0}(\boldsymbol{r},t)}e^{i\phi_{0}(\boldsymbol{r},t)}$,
we get the equations of motion for the phase and density
perturbations~\textcolor{red}{\cite{Eckel:2017uqx}}:
\begin{eqnarray}\label{EckelPhase1}
-\frac{\hbar}{U}\partial_{t}\hat{\phi}_{1} & = &
\hat{\mathcal{D}}\hat{n}_{1} +
\frac{\hbar^{2}}{MU}\boldsymbol{\nabla}\phi_{0}\cdot\boldsymbol{\nabla}\hat{\phi}_{1},
\\ \label{EckelDensity1} \partial_{t}\hat{n}_{1} & = &
-\frac{\hbar}{M}\boldsymbol{\nabla}\cdot\Big[\hat{n}_{1}\boldsymbol{\nabla}\phi_{0}
  + n_{0}\boldsymbol{\nabla}\hat{\phi}_{1} \bigg],
\end{eqnarray}
where we made use of the continuity equation for $n_{0}$: $\partial_{t}n_{0} =
-\frac{\hbar}{M}\boldsymbol{\nabla}\cdot\big(n_{0}\boldsymbol{\nabla}\phi_{0}\big)$ and
we defined the operator
\begin{equation}
  \label{Eq:curdee}
\hat{\mathcal{D}}\equiv
1-\frac{\hbar^{2}}{2MU}\Big(\frac{\nabla^{2}}{2n_{0}} -
\frac{\boldsymbol{\nabla} n_{0}\cdot\boldsymbol{\nabla}}{2n_{0}^{2}} -
\frac{\nabla^{2}n_{0}}{2n_{0}^{2}} + \frac{(\boldsymbol{\nabla}
  n_{0})^{2}}{2n_{0}^{3}}\Big).
\end{equation}
The terms in parentheses in $\hat{\mathcal{D}}$ are due to the quantum pressure, which
are often neglected 
 in the hydrodynamic limit
where $\hat{\mathcal{D}}\approx 1$. In that situation, Eqs.~(\ref{EckelPhase1}) and (\ref{EckelDensity1}) can be readily combined
to get a relativistic wave equation for $\hat{\phi}_{1}$ (see
\textcolor{red}{\cite{Jain:2007gg,Prain:2010zq,Eckel:2017uqx}}) :
\begin{equation}\label{RelativisticWaveEqn}
0 = \frac{1}{\sqrt{-g}}\partial_{\mu}(\sqrt{-g}g^{\mu\nu}\partial_{\nu}\hat{\phi}_{1}),
\end{equation}
where $\mu=0$ denotes time and $\mu=1,2,3$ denote space so
$x^{\mu}=(ct,x,y,z)$ and
$\partial_{\mu}=(\frac{1}{c}\frac{\partial}{\partial
  t},\boldsymbol{\nabla})$ where $c= \sqrt{Un_{0}/M}$ is the BEC speed of sound.
Here, the metric is  
\begin{equation}
\label{Metric}
g_{\mu\nu} = \begin{bmatrix}
-c^{3} & 0 & 0 & 0 \\
0 & c(R+\tilde{\rho})^{2} & 0 & 0 \\
0 & 0 & c & 0 \\
0 & 0 & 0 & c \\
\end{bmatrix},
\end{equation} 
with determinant 
$g=-c^{6}(R+\tilde{\rho})^{2}$, where 
 $\tilde{\rho}=\rho-R(t)$ is the comoving radial
coordinate.  

Equation~(\ref{RelativisticWaveEqn}) shows that a boson gas in a time-dependent
toroidal trap indeed simulates an expanding one-dimensional universe with
the metric Eq.~(\ref{Metric}).  However, below we show that although
the quantum pressure terms in $\hat{\mathcal{D}}$ are small, their inclusion
qualitatively impacts the dynamics of low-energy modes in the expanding
toroidal BEC, leading to damping and spontaneous phonon creation. 
  This quantum pressure provides short distance corrections to the evolution of sound
  modes in a BEC, which become significant when we reduce the thickness of the ring to make it quasi one-dimensional.

Having obtained equations
(\ref{EckelPhase1}) and (\ref{EckelDensity1}), that describe
excitations of a superfluid boson gas in a time-dependent trap, in the
next section we show how, in the thin-ring limit, these equations
reduce to the Mukhanov-Sasaki equation that describes damped sound
modes.

\section{The Mukhanov-Sasaki Equation}
\label{sec:MSE}
Equations~(\ref{EckelPhase1}) and (\ref{EckelDensity1}) derived in the preceding section
describe phase and density perturbations (i.e., phonons) in a BEC with a generic time-dependent trapping potential
$V(\boldsymbol{r},t)$.
In fact, the potential $V(\boldsymbol{r},t)$ only explicitly appears in the dynamics of the background density
($n_0$) and phase ($\phi_0$) on which phonons propagate (which we study in Appendix~\ref{appendix}), while
the density and phase perturbations $\hat{n}_1$ and $\hat{\phi}_1$ are sensitive to $n_0$ and $\phi_0$
via Eqs.~(\ref{EckelPhase1}) and (\ref{EckelDensity1}).
Our first task is to make simplifying approximations that apply
to the geometry realized in Ref.~\onlinecite{Eckel:2017uqx}, i.e., a thin expanding toroidal trapping potential.
As we shall see, this leads to the Mukhanov-Sasaki equation for damped sound modes.

The first simplifying approximation we shall invoke is to neglect
the $\rho$ and $z$ dependences of the phase and density perturbations,
thereby replacing $\hat{\phi}_1(\br,t) \to \hat{\phi}_1(\theta,t)$ and
$\hat{n}_1(\br,t) \to \curV^{-1} \hat{n}_1(\theta,t)$ in Eqs.~(\ref{EckelPhase1}) and (\ref{EckelDensity1}),
where $\curV = R\curA$ is a volume scale  with $\curA$ the cross-sectional area of the toroid (so
that $2\pi\curV$ is the toroid volume).  We note that 
the angle-dependent density fluctation operator $\hat{n}_1(\theta)$ is dimensionless.

Such an approximation holds in the thin-ring limit, where the toroidal radius $R(t)$
is much larger than the typical length scales  $R_\rho$ and $R_z$  (defined in Appendix~\ref{appendix})
characterizing the ring cross-sectional area (see Fig.~\ref{expandingSketch}).
This implies that an initial angle-dependent perturbation
around the ring, such as prepared in the experiments of Ref.~\textcolor{red}{\cite{Eckel:2017uqx}},
will not excite density variations in the $\rho$ and $z$ directions.

The second simplifying approximation we shall invoke is to assume that 
the background phase $\phi_{0}$ and density $n_{0}$ are functions
only of
$\rho$ and $z$ (i.e., they are independent of $\theta$).  This is
expected, given  the angular symmetry of the toroidal trapping potential.  
We shall furthermore assume that the condensate and the ring are moving
with the same velocity, i.e., the superfluid velocity equals the ring velocity
$\boldsymbol{v}=\frac{\hbar}{M}\boldsymbol{\nabla}\phi_{0}=\dot{R}\hat{\rho}$ (here $\dot{R} \equiv \frac{dR}{dt}$).
The conditions for validity of this assumption are explored in Appendix~\ref{appendix}. 
This implies that the gradients in the perturbations are
orthogonal to the condensate velocity
i.e. $\boldsymbol{v}\cdot\boldsymbol{\nabla}\hat{\phi}_{1} =
\boldsymbol{v}\cdot\boldsymbol{\nabla}\hat{n}_{1} = 0$. By a similar
argument, the dot-product term $\boldsymbol{\nabla}
n_{0}\cdot\boldsymbol{\nabla} \hat{n}_{1}=0$ in the quantum pressure
also vanishes. On the other hand the divergence of condensate velocity
is not zero:
$\boldsymbol{\nabla}\cdot\boldsymbol{v}=\frac{\hbar}{M}\nabla^{2}\phi_{0}\approx\frac{\dot{R}}{R}$.

Within these approximations, 
the equations of motion (\ref{EckelPhase1}) and (\ref{EckelDensity1})
in the thin ring limit
take the form:
\begin{eqnarray}\label{EckelPhase}
-\frac{\hbar\curV}{U}\partial_{t}\hat{\phi}_{1}(\theta,t) &=& 
\hat{\mathcal{D}}_{\theta}\hat{n}_{1}(\theta,t), \\ \label{EckelDensity}
\partial_{t}\hat{n}_{1}(\theta,t)
 &=& -\frac{\dot{R}}{R}\hat{n}_{1}(\theta,t)
 - \frac{\hbar \curV n_{0}}{MR^{2}}\partial_{\theta}^{2}\hat{\phi}_{1}(\theta,t),~~~
\end{eqnarray}
describing angle-dependent excitations in a thin radially expanding toroidal BEC.
Here $\hat{\mathcal{D}}_{\theta}\equiv 1-\frac{\hbar^{2}}{2MU}\Big(\frac{\partial_{\theta}^{2}}{2n_{0}R^{2}} - \frac{\nabla^{2}n_{0}}{2n_{0}^{2}} + \frac{(\boldsymbol{\nabla} n_{0})^{2}}{2n_{0}^{3}}\Big)$ is the projection of $\hat{\mathcal{D}}$ in the $\theta$-space.
To solve this system of equations, we introduce mode expansions as :
\begin{eqnarray}\label{ModeExpPhi1QP}
&&\hspace{-1cm}\hat{\phi}_{1}(\theta,t) \!= \!\sqrt{\frac{U}{2\pi\curV\hbar}}
\!\sum_{n=-\infty}^{\infty}\!\!\bigg[e^{in\theta}\chi_{n}(t)\hat{a}_{n}\!+\!e^{-in\theta}\chi_{n}^{*}(t)\hat{a}^{\dagger}_{n}\bigg], 
\\ \label{ModeExpn1QP}
&&\hspace{-1cm}\hat{n}_{1}(\theta,t) \!= \!\sqrt{\frac{ U\curV}{2\pi\hbar}}
\sum_{n=-\infty}^{\infty}\!\!\bigg[e^{in\theta}\eta_{n}(t)\hat{a}_{n}\!+\!e^{-in\theta}\eta_{n}^{*}(t)\hat{a}^{\dagger}_{n}\bigg], 
\end{eqnarray} 
where
the ladder operators
$\hat{a}_{n} $ satisfy
$\big[\hat{a}_{n},\hat{a}_{n'}^{\dagger}\big]=\delta_{n,n'}$, and the
mode functions are assumed to be same whether the modes are traveling
clockwise or anticlockwise, i.e. $\chi_{-n}=\chi_{n}$ and
$\eta_{-n}=\eta_{n}$.  We take $\chi_{n}(t)$ and $\eta_{n}(t)$ to satisfy $(\eta_{n}\chi^{*}_{n}-\eta^{*}_{-n}\chi_{-n}) = i\hbar/U$, which leads to the
commutation relation 
$\big[\hat{n}_{1}(\theta,t),\hat{\phi}_{1}(\theta',t)\big] =
i\delta(\theta-\theta')$.
Substituting these mode expansions in (\ref{EckelPhase}) and
(\ref{EckelDensity}), we get the following equations of motion
for the mode functions:
\begin{eqnarray}\label{RelationDensityPhase}
-\frac{\hbar}{U}\dot{\chi}_{n}(t) & = & D_{n}\eta_{n}(t), \\
\dot{\eta}_{n}(t) & = & -\frac{\dot{R}}{R}\eta_{n} + \frac{\hbar n_{0}}{MR^{2}}n^{2}\chi_{n},
\end{eqnarray}
where $D_{n} \equiv 1 + \frac{\hbar^{2}}{4M^{2}c^{2}}\Big(\frac{n^{2}}{R^{2}} + \frac{\nabla^{2}n_{0}}{n_{0}} - \frac{(\nabla n_{0})^{2}}{n_{0}^{2}}\Big)$ is the eigenvalue of $\hat{\mathcal{D}}_{\theta}$. 

To arrive at the Mukhanov-Sasaki equation, we eliminate $\eta_{n}$ in favor of $\chi_{n}$:
\begin{equation}\label{MukhanovSasakiQP}
\ddot{\chi}_{n} + \big(1+\gamma_{\text{QP}}\big)\frac{\dot{R}}{R}\dot{\chi}_{n} + \alpha_{\text{QP}}\frac{n^{2}c^{2}}{R^{2}}\chi_{n} = 0.
\end{equation}
The corrections due to quantum pressure are
$\gamma_{\text{QP}}=-\frac{R}{\dot{R}}\cdot\frac{\dot{D}_{n}}{D_{n}}$
and $\alpha_{\text{QP}}=D_{n}$. In general these are dependent on
the density, the radius of the ring and the mode index $n$. However, we can make
an estimate of what values these corrections typically take. We
start by approximating the density gradient
$\boldsymbol{\nabla}n_{0}$
by the density divided by
the width of
the ring $w$ (characterized by the TF radii $R_\rho$ and $R_z$, see
Appendix~\ref{appendix}).
Similarly, we can approximate $\nabla^2 n_0\simeq -n_0/w^2$.  Thus
$D_{n}\approx
1+\frac{n^{2}}{2}\big(\frac{\xi}{R}\big)^{2}-\big(\frac{\xi}{w}\big)^{2}$,
where $\xi=\frac{\hbar}{\sqrt{2}Mc}$ is the coherence length.
In the
hydrodynamic limit, where the coherence length is small compared to
the dimensions of the ring, $\alpha_{QP} = D_n \approx 1$.

To estimate $\gamma_{\text{QP}}$, we need to estimate $\dot{D}_{n}$.  To do this
we use experimental parameters from
Ref.~\textcolor{red}{\onlinecite{Eckel:2017uqx}}, with $M$ given by
the mass of a $^{23}\text{Na}$ atom, the speed of sound $c\approx
2~\text{mm/s}$ and the width of the ring $w\approx 2~\mu\text{m}$
(which is indeed an order of magnitude smaller than the radius $R(t)$
that varies between $10~\mu\text{m}$ to
$50~\mu\text{m}$~\textcolor{red}{\cite{Eckel:2017uqx}}). Thus
$\big(\frac{\xi}{w}\big)^{2} \approx 10^{-1}$. Also, since the width
of the ring is small compared to its radius, $w\ll R(t)$, we find
$\dot{D}_{n}\approx
-\frac{1}{2}\big(\frac{\xi}{w}\big)^{2}\frac{\dot{R}}{R}$. Here we
estimate the ring width via $w=\sqrt{\frac{2\mu}{M\omega_{z}^{2}}}$,
where $\omega_{z}$ is the trapping frequency in the $z$ direction. We
also made use of the local density approximation (LDA) to estimate the
chemical potential in a harmonic
trap. Following~\textcolor{red}{\cite{Eckel:2017uqx}}, we get
$\mu\propto R^{-1/2}$ (see Appendix~\ref{appendix}). This implies that
$\gamma_{\text{QP}} \approx
\frac{1}{2}\big(\frac{\xi}{w}\big)^{2}\approx 10^{-1}$. 
  This shows that short-distance physics of quantum pressure comes into play as we
  decrease the width of the ring $w$ to make it quasi 1D.

\begin{figure}[ht!]
   \begin{center}
     \includegraphics[width=0.5\textwidth]{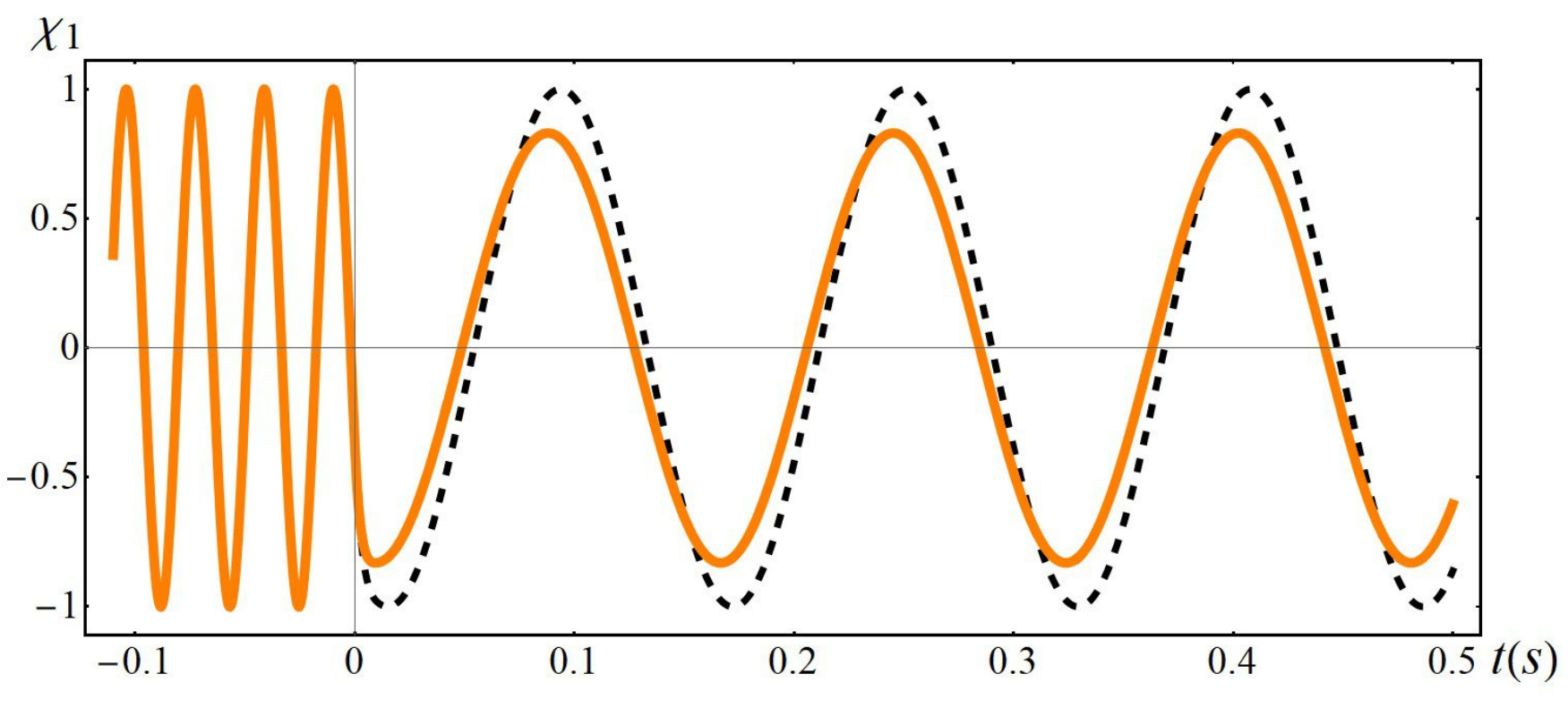}
     \end{center}
\caption{(Color Online) Comparison of solutions $\chi_{m}$ of (\ref{MukhanovSasaki})
for $m=1$ mode without quantum pressure ($\gamma=0$) ({\em dashed, black})
and with quantum pressure ($\gamma=0.5$) ({\em solid, orange}) (for $t<0$ they coincide).
The modes undergo
damping and redshift just as they do in the inflationary era.  For this plot, we took the speed of sound to be $c=2~$mm/s, the initial radius to be $R_{0}=10\mu$m, the timescale that governs the expansion of the ring to be  $\tau=6.21\text{ms}$ and the duration of expansion to be $t_{f}=10$ms.}
\label{damping}
\end{figure}

We pause to note that although the quantum-pressure terms in Eq.~(\ref{Eq:curdee}) do not have a
{\em direct\/} cosmological analogue, our final Mukhanov-Sasaki equation  is
in fact relevant for cosmology.  Indeed, within inflationary cosmology,
the Einstein equations that describe the background evolution of inflationary spacetime
yield a  Mukhanov-Sasaki equation that is of the form of Eq.~(\ref{MukhanovSasakiQP})
but with the coefficient of the damping term being the number of spatial dimensions,
with a small correction due to the so-called ``slow-roll parameters''~\textcolor{red}{\cite{Weinberg:2008zzc}}
that is reminscent of the small parameter $\gamma_{\text{QP}}$ obtained here.  More generally, as we shall show below,
Eq.~(\ref{MukhanovSasakiQP}) leads to a crucial prediction of inflationary theory, which is
spontaneous particle creation that leads (in the cosmological context) to the  distribution of anisotropies in the CMB.

In fact, the quantum pressure effects in Eq.~(\ref{MukhanovSasakiQP}) are essential
for achieving particle production.  To see this, we now examine the implications of making
a futher approximation (motivated by our preceding estimates),
 in which we take 
$\alpha_{\text{QP}} \to 1$
and $\gamma_{\text{QP}} \to 0$.  Indeed, this approximation brings
Eq.~(\ref{MukhanovSasakiQP}) into a very simple form:
\begin{eqnarray}\label{MSeasy}
\ddot{\chi}_{n} + \frac{\dot{R}}{R}\dot{\chi}_{n} +
\frac{n^{2}c^{2}}{R^{2}}\chi_{n} = 0,
\end{eqnarray}
where we assume a constant speed of sound $c$. Equation~(\ref{MSeasy})
was discussed
in detail in~\textcolor{red}{\cite{Eckel:2017uqx}}.
Here, we solve it by introducing the 
conformal time $d\eta=\frac{cdt}{R(t)}$ that measures the size
in cosmology. Equation (\ref{MSeasy}) then takes the form of a simple
harmonic oscillator: $\chi_{n}''(\eta)+n^{2}\chi(\eta)=0$, with plane
waves $e^{\pm i|n|\eta}$ as solutions. Switching back to proper
time $t$, we get:
\begin{eqnarray}
  \label{Eq:chiproper}
\chi_{n}(t)\Big|_{\gamma=0} \sim \exp\bigg[\pm
  i|n|\int_{0}^{t}\frac{cdt'}{R(t')}\bigg].
\end{eqnarray}
This shows that the amplitude of the modes will be conserved with time
in the $\gamma_{\text{QP}} \to 0$ limit. Thus the quantum pressure
correction $\gamma_{\text{QP}}$ plays the role of a damping
parameter. As we will see in Sec. \ref{sec:SPC}, nonzero
$\gamma_{\text{QP}}$ (even if it is small in magnitude) is essential
for particle production.  This is a well-known fact in cosmology
\textcolor{red}{\cite{Fulling:1989nb}}, where a conformally invariant
scalar field living in a $(1+1)$-dimensional spatially flat FLRW
spacetime, has plane wave modes as solutions of the wave equation and
thus leads to no particle creation.

Thus, while small, the quantum pressure correction represented by
$\gamma_{\text{QP}}$ fundamentally modifies the solutions of the
Mukhanov-Sasaki equation. In general, the parameters
$\gamma_{\text{QP}}$ and $\alpha_{\text{QP}}$ are dependent on time,
but in what follows, we will assume them to be constants
($\gamma_{\text{QP}}=\gamma$, $\alpha_{\text{QP}} = \alpha$). This approximation gives us an analytic
handle on the parameter space, where the basic physical features like
the particle creation can be modeled without going in to the fine
details of how they 
evolve with time.
  We will take $\gamma$ to be a small number and 
  take $\alpha\to 1$. The reason for the latter approximation is that in this paper, we are mainly
  interested in particle creation due to quantum pressure, embodied in the parameter $\gamma$.  We expect that the slight
  deviation of $\alpha$ from unity, which effectively yields a time-dependent speed of sound (see \textcolor{red}{\cite{Fedichev:2003id}},\textcolor{red}{\cite{Barcelo:2003wu,Fedichev:2003bv,Fischer:2004bf,Jain:2007gg,Prain:2010zq}}), will be a subleading
  effect here.
    Within these approximations,
(\ref{MukhanovSasakiQP}) reduces to the following Mukhanov-Sasaki
equation~\textcolor{red}{\cite{Sasaki:1983kd,Kodama:1985bj,Mukhanov:1988jd}}:
\begin{equation}\label{MukhanovSasaki}
\ddot{\chi}_{n} + \big(1+\gamma\big)\frac{\dot{R}}{R}\dot{\chi}_{n} + \frac{n^{2}c^{2}}{R^{2}}\chi_{n} = 0,
\end{equation}
where we have assumed the speed of  sound $c$ to be a constant.

In the following, we also choose a specific form for the time-dependent radius $R(t)$.
Our choice is motivated by the fact that,  in the inflationary era, the Hubble parameter
$H\equiv\frac{\dot{a}}{a}\approx H_{0}$ is roughly a constant and one
could model this phase with a de-Sitter type inflation where the scale
factor $a(t)\sim e^{H_{0}t}$~\cite{Peebles:1994xt,Weinberg:2008zzc}. This motivates us to study an exponential
expansion $R(t)=R_{0}e^{t/\tau}$ of the ring radius, characterized by the timescale $\tau$. Then the
general solution to Eq.~(\ref{MukhanovSasaki}) is:
\begin{equation}\label{DampingSolution}
\chi_{n}(t) = e^{-\frac{t}{2\tau}(1+\gamma)}\Big[A_{n}J_{\frac{1+\gamma}{2}}(z) + B_{n}J_{-\frac{1+\gamma}{2}}(z)\Big],
\end{equation}
where the time dependent parameter $z=\omega_{n}\tau$ with the
frequency $\omega_{n}=\frac{|n|c}{R(t)}$ 
and $J_{n}(z)$ are Bessel functions of the first kind. The
coefficients $A_{n}$ and $B_{n}$ are fixed by the initial conditions.

To illustrate the effect of nonzero
$\gamma$ in the case of exponential inflation in the toroid, in
Fig.~\ref{damping} we plot $\chi_{n}$  as a function of $t$ for
an $n=1$ mode for the case of $\gamma=0$
(dashed curve) and $\gamma =0.5$ (solid curve).  For $t<0$ (before expansion), both curves show
oscillatory motion, and during expansion (for $t>0$), where they are governed by Eq.~(\ref{DampingSolution}),
they both show a redshift~\textcolor{red}{\cite{Eckel:2017uqx}} i.e. their frequency
$\omega_{n}=\frac{nc}{R(t)}$ decreases as the ring expands.  However, while the 
$\gamma=0$ curve shows no reduction of amplitude, the $\gamma =0.5$  curve exhibits
damping due to quantum pressure. 
Both of these phenomena, the redshift and damping, have their 
respective counterparts in cosmology.

Before concluding this section, we note that, as in conventional inflationary theory,
the fate of modes after expansion in a toroidal BEC is strongly dependent on the mode index $n$ (which controls
the mode wavelength).
To see this, note that
for exponential expansion $R(t) \propto e^{t/\tau}$, the horizon size is given
by
$\eta_{\rm H}=\int_{0}^{t_{f}}\frac{cdt}{R(t)}=c\tau(1-e^{-t_{f}/\tau})$.
After
the expansion persists long enough i.e. $t_{f}/\tau$ is large, then
the horizon size is 
$\eta_{\rm H} \sim c\tau$. If the parameter
$z=\frac{|n|c\tau}{R(t)}$ is small, i.e., $\frac{R}{|n|}\gg c\tau$, then the 
wavelengths of the modes are much larger than the horizon size and
the mode solution (\ref{DampingSolution}) becomes constant in time:
\begin{eqnarray}\label{Freezing}
  \chi_{n}(t) & \approx &
  \bigg(\frac{2R_0}{|n|c\tau}\bigg)^{\frac{1+\gamma}{2}} \frac{B_{n}}
      {\Gamma\big(\frac{1-\gamma}{2}\big)},
\end{eqnarray}
where we have used the result that for small arguments $z\rightarrow
0$, the Bessel function goes as
$J_{n}(z)\rightarrow\frac{1}{\Gamma(n+1)}\big(\frac{z}{2}\big)^{n}$.
  Note that the form of the exponential factor in Eq. (\ref{DampingSolution}) is
  essential to get the above result.
This
{\em freezing\/} of super-horizon modes (i.e. those with small mode index) is a very
important aspect of the inflationary mechanism as these modes re-enter
the horizon at a later time and form large scale structures in the
observable universe. In the next section, we will discuss how phonons
are produced due to the mode solution (\ref{DampingSolution}) and see
the importance of the super-horizon modes in the expanding ring.

\section{Spontaneous Phonon Creation}
\label{sec:SPC}
Now that we have solved the Mukhanov-Sasaki equation for a constant
quantum pressure parameter $\gamma$, in this section we use its solution
(\ref{DampingSolution}) to understand how a BEC in the vacuum state,
when expanded exponentially, will exhibit the dynamical generation of
phonons.  We start with an initially static BEC in its vacuum state.
The mode functions for this initial BEC (which we denote as
the ``in'' state)  obey Eq.~(\ref{MukhanovSasaki}), but with $\dot{R} = 0$
(so that they are undamped).  
These inital mode functions are:
\begin{equation}\label{InModes}
\chi^{\text{in}}_{n}(t) = \frac{1}{\sqrt{2\omega^{0}_{n}}}e^{-i\omega^{0}_{n}t},
\end{equation}
which satisfy
$i\partial_{t}\chi^{\text{in}}_{n}=\omega^{0}_{n}\chi^{\text{in}}_{n}$,
where $\omega^{0}_{n}$ is the frequency at $t=0$. These
positive-frequency \lq in\rq-mode functions satisfy the Wronskian condition
$W[\chi_{n},\chi^{*}_{n}]=\dot{\chi}_{n}\chi^{*}_{n}-\chi_{n}\dot{\chi}^{*}_{n}=-i$,
and associated with them is the structure of ladder operators
$\hat{a}_{n}$ that annihilate their associated `a-vacuum' state:
$\hat{a}_{n}|0_{a}\rangle=0$.

Having described the normal modes of the initial BEC, we turn to the
impact of a period of exponential growth on the BEC, starting at $t=0$,
and described by the exponential function
$R(t)=R_{0}e^{t/\tau}$. During this period, the modes evolve according
to (\ref{DampingSolution}), where the coefficients are fixed by
matching the mode functions $\chi_n(t)$ and their time derivatives with that of the initial BEC at $t=0$.
This matching results in the conditions:
\begin{eqnarray}\label{AnBn}
A_{n} & = & \frac{J_{\frac{1-\gamma}{2}}(z_{0})+iJ_{-\frac{1+\gamma}{2}}(z_{0})}{J_{\frac{1+\gamma}{2}}(z_{0})J_{\frac{1-\gamma}{2}}(z_{0})+J_{-\frac{1+\gamma}{2}}(z_{0})J_{\frac{-1+\gamma}{2}}(z_{0})},~ \nonumber \\
B_{n} & = & \frac{J_{\frac{-1+\gamma}{2}}(z_{0})-iJ_{\frac{1+\gamma}{2}}(z_{0})}{J_{\frac{1+\gamma}{2}}(z_{0})J_{\frac{1-\gamma}{2}}(z_{0})+J_{-\frac{1+\gamma}{2}}(z_{0})J_{\frac{-1+\gamma}{2}}(z_{0})},~~~~~
\end{eqnarray}
where $z_{0}=\frac{|n|c\tau}{R_{0}}$.  Equation~(\ref{DampingSolution}), along with these coefficients, describes the mode functions during the exponential
growth regime of the toroidal BEC. 

\begin{figure}[h]
\centering
\includegraphics[width=0.5\textwidth]{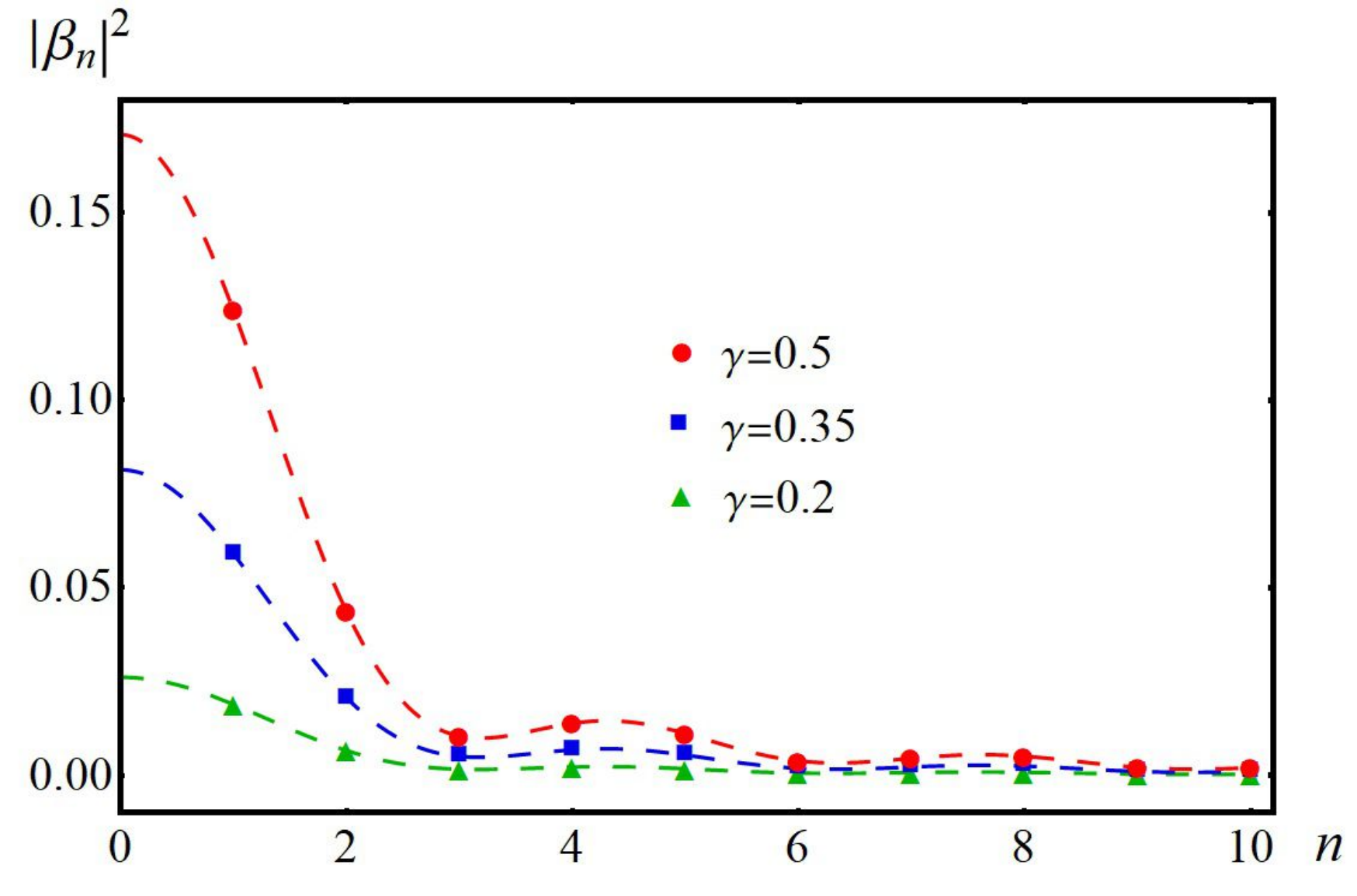}
\caption{(Color Online) The particle production parameter $|\beta_{n}|^{2}$ as function of mode index $n$ for various values
  of the quantum pressure $\gamma=0.2$ ({\em green\/}),
  $\gamma=0.35$ ({\em blue\/}) and $\gamma=0.5$ ({\em red\/}). The dashed lines
  show  $|\beta_{n}|^{2}$ as continuous functions to emphasize their overall dependence on the discrete mode indices. For this plot, we took the speed of sound to be $c=2~$mm/s, the initial radius to be $R_{0}=10\mu$m, the expansion timescale to be  $\tau=6.21\text{ms}$ and the duration of expansion to be $t_{f}=10$ms.
}
\label{Beta}
\end{figure}

The expansion comes to a halt at some later time $t=t_{f}$.  In this third regime of $t>t_f$,
the BEC Hamiltonian is again static, with excitations described by \lq out\rq-mode solutions that
are analogous to Eq.~(\ref{InModes}):
\begin{equation}\label{OutModes}
\chi^{\text{out}}_{n}(t) = \frac{1}{\sqrt{2\omega^{f}_{n}}}e^{-i\omega^{f}_{n}(t-t_{f})}.
\end{equation}
Due to the quantum evolution during the ring expansion, the final Heisenberg picture results for the operators
$\hat{\phi}_{1}(\theta,t)$ and $\hat{n}_{1}(\theta,t)$ are still given by Eqs.~(\ref{ModeExpPhi1QP}) but with
the final mode functions a superposition of the `out'-modes in Eq.~(\ref{OutModes}):
\begin{equation}\label{NewMode}
\chi^{f}_{n}(t) = e^{-\frac{t_{f}}{2\tau}\gamma}\Big[\alpha_{n}\chi^{\text{out}}_{n}(t) + \beta_{n}\chi^{\text{out}*}_{n}(t)\Big],
\end{equation}
where $\omega^{f}_{n}=\frac{|n|c}{R_{f}}$ is the frequency for $t\geq t_{f}$.  The coefficients in Eq.~(\ref{NewMode}) are obtained by again demanding that
the mode function and its derivatives are consistent at $t_f$, with the $t\to t_f^-$ solution given by Eq.~(\ref{DampingSolution}) as described above.  By
solving these matching conditions we find the coefficients $\alpha_n$ and $\beta_n$: 
\begin{eqnarray}\label{alpha-n}
&&\hspace{-1cm}  \alpha_{n}\! \!=\!\! \frac{e^{-\frac{t_{f}}{2\tau}}}{2}\Big[\! A_{n}\Big(J_{\frac{1+\gamma}{2}}\!-\!iJ_{\frac{-1+\gamma}{2}}\Big)
    \!+\! B_{n}\Big(J_{-\frac{1+\gamma}{2}}\!+\!iJ_{\frac{1-\gamma}{2}}\Big)\!\Big],  \\ \label{beta-n}
&&\hspace{-1cm}  \beta_{n}\!\! = \!\! \frac{e^{-\frac{t_{f}}{2\tau}}}{2}\Big[\! A_{n}\Big(J_{\frac{1+\gamma}{2}}\!+\!iJ_{\frac{-1+\gamma}{2}}\Big)
    \!+\! B_{n}\Big(J_{-\frac{1+\gamma}{2}}\!-\!iJ_{\frac{1-\gamma}{2}}\Big)\!\Big], 
\end{eqnarray}
where we have suppressed the arguments of the  Bessel functions, which
are all evaluated at $z_{f}=\frac{|n|c\tau}{R_{f}}$ with $R_{f}$  the
final ring radius.  These coefficients, which satisfy
$|\alpha_{n}|^{2}-|\beta_{n}|^{2}=1$, describe the modification of the
mode functions $\chi_n$ during the expansion process.
  One can plug in equations (\ref{alpha-n})-(\ref{beta-n}) in
  Eq. (\ref{NewMode}), and thereby realize that the final mode
  $\chi^{f}_{n}$ has the same form as Eq. (\ref{DampingSolution}),
  i.e. an exponentially decreasing factor times some linear
  combination of Bessel functions.

The modified mode function (\ref{OutModes}) in the `out'-regime,
defines a new set of ladder operators $\hat{b}_{n}$ that annihilate
the new `b-vacuum' state $|0_{b}\rangle\neq |0_{a}\rangle$:
$\hat{b}_{n}|0_{b}\rangle=0$. The coefficients $\alpha_{n}$ and
$\beta_{n}$ provide a Bogoliubov transformation between $\hat{a}_{n}$
and $\hat{b}_{n}$
via~\textcolor{red}{\cite{Birrell:1982ix,Mukhanov:2007zz,Parker and Toms,Fulling:1989nb}}:
\begin{eqnarray}\label{BogoliuobovTransformation}
\hat{a}_{n} & = & \alpha^{*}_{n}\hat{b}_{n}-\beta^{*}_{n}\hat{b}^{\dagger}_{-n}, \\ \label{InverseBogoliubov}
\hat{b}_{n} & = & \alpha_{n}\hat{a}_{n}+\beta^{*}_{n}\hat{a}^{\dagger}_{-n}.
\end{eqnarray}
Thus, assume we start in the `a-vacuum', characterized by vanishing particle density
$\langle 0_{a}|\hat{n}^{a}|0_{a}\rangle=0$, where $\hat{n}^{a}=\hat{a}^{\dagger}_{n}\hat{a}_{n}$.
During the expansion, 
the system wavefunction remains $|0_{a}\rangle$
(since we work in the Heisenberg picture),
but the  $\hat{a}_n$ evolve into the
$\hat{b}_n$ according to Eq.~(\ref{BogoliuobovTransformation}).
A measurement of
the particle density will find the final system is bubbling with `b-particles'~\cite{Carroll:2004st},
as represented by the expectation value $\langle 0_{a}|\hat{n}^{b}|0_{a}\rangle=|\beta_{n}|^{2}$.
This is known as
spontaneous particle creation from the vacuum state, characterized by
the power parameter $|\beta_{n}|^{2}$ that we plot in Fig.~\ref{Beta}
with respect to the mode index $n$, for three values of the quantum
pressure parameter $\gamma=0.2$, $\gamma=0.35$ and $\gamma=0.5$.

We now describe the connection of these results to the theory of inflation.
We  can infer from Fig.~\ref{Beta} that the power associated with small mode
indices such as $n=1$ is much higher than those at larger $n$. As in
inflationary cosmology, this is because
during the expansion some modes such as $n=1$ become super-horizon and
freeze (\ref{Freezing}), i.e. their power remains constant. In contrast,
modes that are well within the horizon (i.e., at higher $n$, or smaller wavelength)
are strongly damped and thus their power $|\beta_{n}|^{2}$ is reduced.

Thus, larger values of the quantum pressure parameter lead to stronger damping of
modes (\ref{MukhanovSasaki}) as well as increased spontaneous phonon production.  This suggests that
the loss of amplitude is converted into  phonon production.
In the next section, we will discuss the possibility of observing
stimulated creation of phonons from a coherent initial state.

\section{Stimulated Phonon Creation}
\label{sec:STIM}
Having discussed how phonons can be produced by a BEC that is initially in its vacuum state, we now turn to the possibility of starting with a initial coherent state in the mode $N$:
\begin{eqnarray}
  \label{eq:systemwave}
|\alpha, N \rangle & = & e^{-\frac{1}{2}|\alpha|^{2}}e^{\alpha \hat{a}^{\dagger}_{N}} |0_{a} \rangle,
\end{eqnarray}
where the complex parameter $\alpha$ is a measure of the average number of particles in the coherent state given by $|\alpha|^{2}$. Physically, such a state represents a macroscopic current-carrying state of the BEC.
\begin{figure}[h]
\centering
\includegraphics[width=0.5\textwidth]{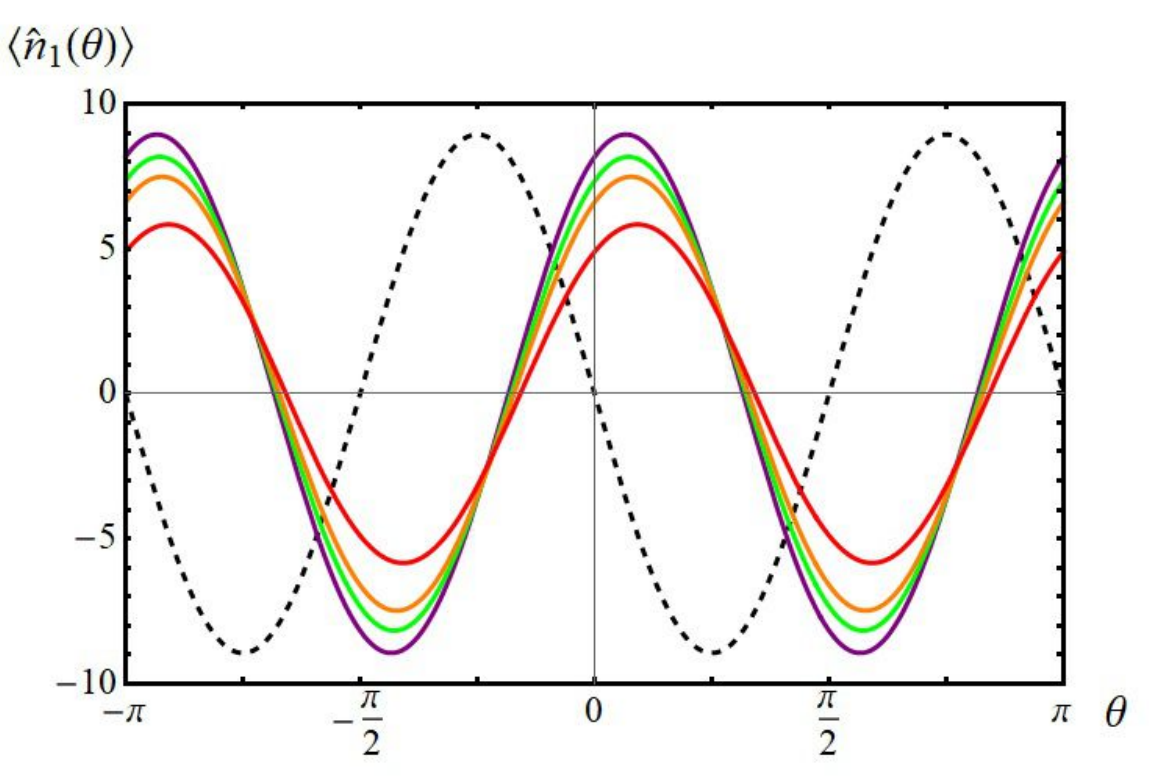}
\caption{(Color Online) Comparison of average density 
  $\langle\hat{n}_{1}\rangle$
  before (\emph{dashed, black\/}) and just after expansion ({\em solid\/}) for
  four values of the quantum pressure parameter $\gamma$.  In order of
  decreasing amplitude, the solid curves correspond to $\gamma=0$
  (\emph{violet}),  $\gamma=0.1$ (\emph{green}),
  $\gamma=0.2$ (\emph{orange}), and $\gamma=0.5$ (\emph{red}).
  Note the $\gamma=0$ final curve has equal amplitude to the
  initial density profile   (in agreement with
  our earlier finding shown in Fig.~\ref{damping}), with the phase
  difference $\alpha_{n}=e^{i(z_{f}-z_{0})}$ between the curves reflecting the fact that the wave
  travels around the ring during expansion. 
For this plot we studied a coherent state characterized by $\alpha=\sqrt{10}$ and mode index $N=2$. We also took the sound speed to be $c=2~$mm/s, the initial radius to be $R_{0}=10\mu$m, the timescale for expansion to be  $\tau=6.21\text{ms}$ and the duration of expansion to be $t_{f}=10$ms.}
\label{density}
\end{figure}
The states $|\alpha, N \rangle$ are eigenfunctions of the annihilation operators:
\begin{eqnarray}
\hat{a}_{m}|\alpha, N \rangle & = & \delta_{m,N}\alpha|\alpha, N \rangle.
\end{eqnarray}
In what follows, we will use these coherent states to calculate 
the average density in the ring,
before and after expansion. The advantage of using
coherent states relative to fixed-number Fock states is that in the
latter, the average density is zero at all times. This implies that,
in an experiment, no significant change will be observed in the average
density.  For the case of an initial vacuum state, studied in the preceding
section, another complication is the presence of other phonon modes,
such as thermally excited phonons (since experiments cannot truly reach
the zero-temperature vacuum state), that may swamp the signal from spontaneously created phonons.  
In contrast, in a coherent state, the average densities
change with time, making it easier to detect changes due to phonon production.

\begin{figure*}[t]
\centering
\subfloat[Subfigure 1 list of figures text][$\Delta t=1.6\tau$]{
\includegraphics[width=0.24\textwidth]{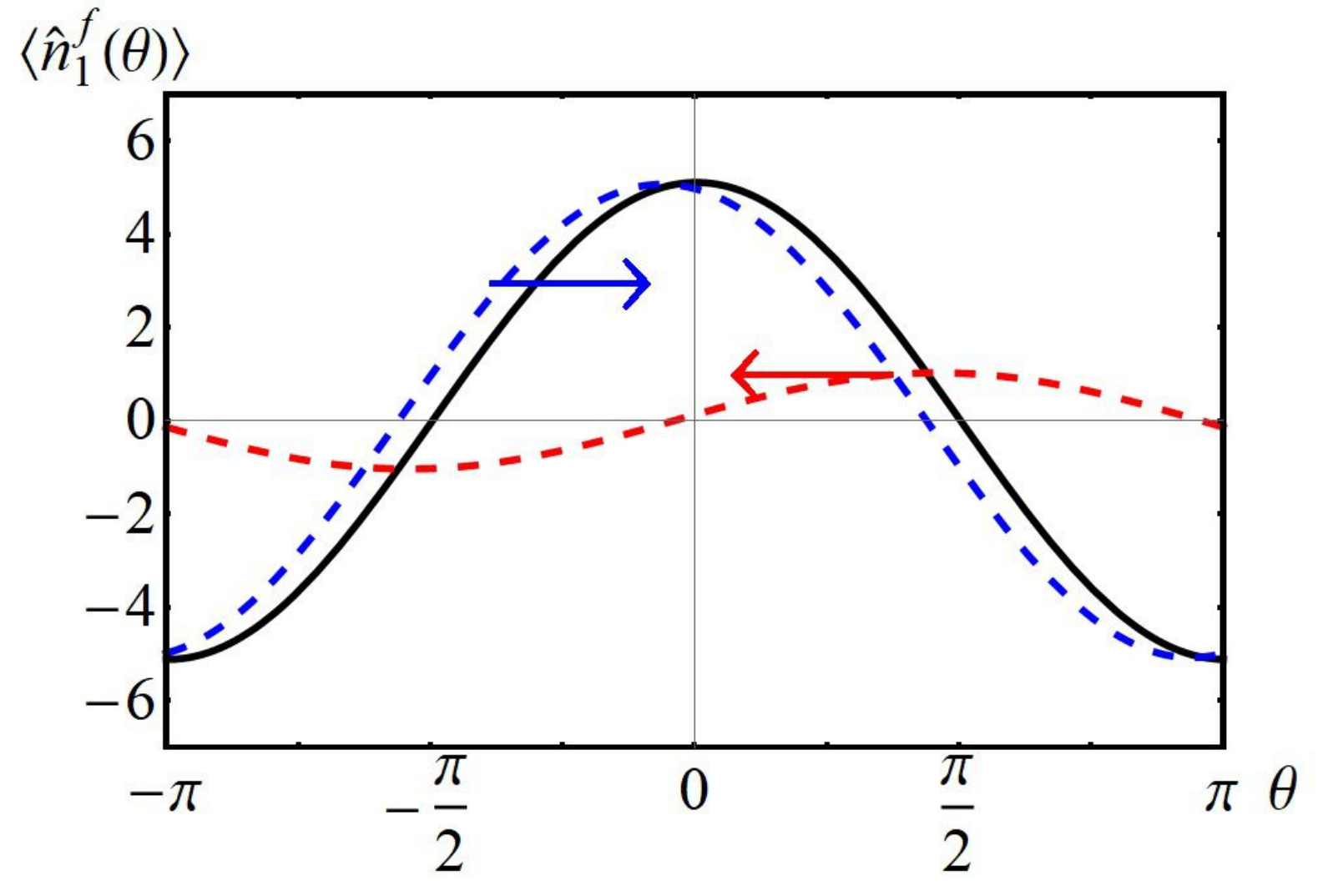}
\label{fig:subfig10}}
\subfloat[Subfigure 2 list of figures text][$\Delta t=4.8\tau$]{
\includegraphics[width=0.24\textwidth]{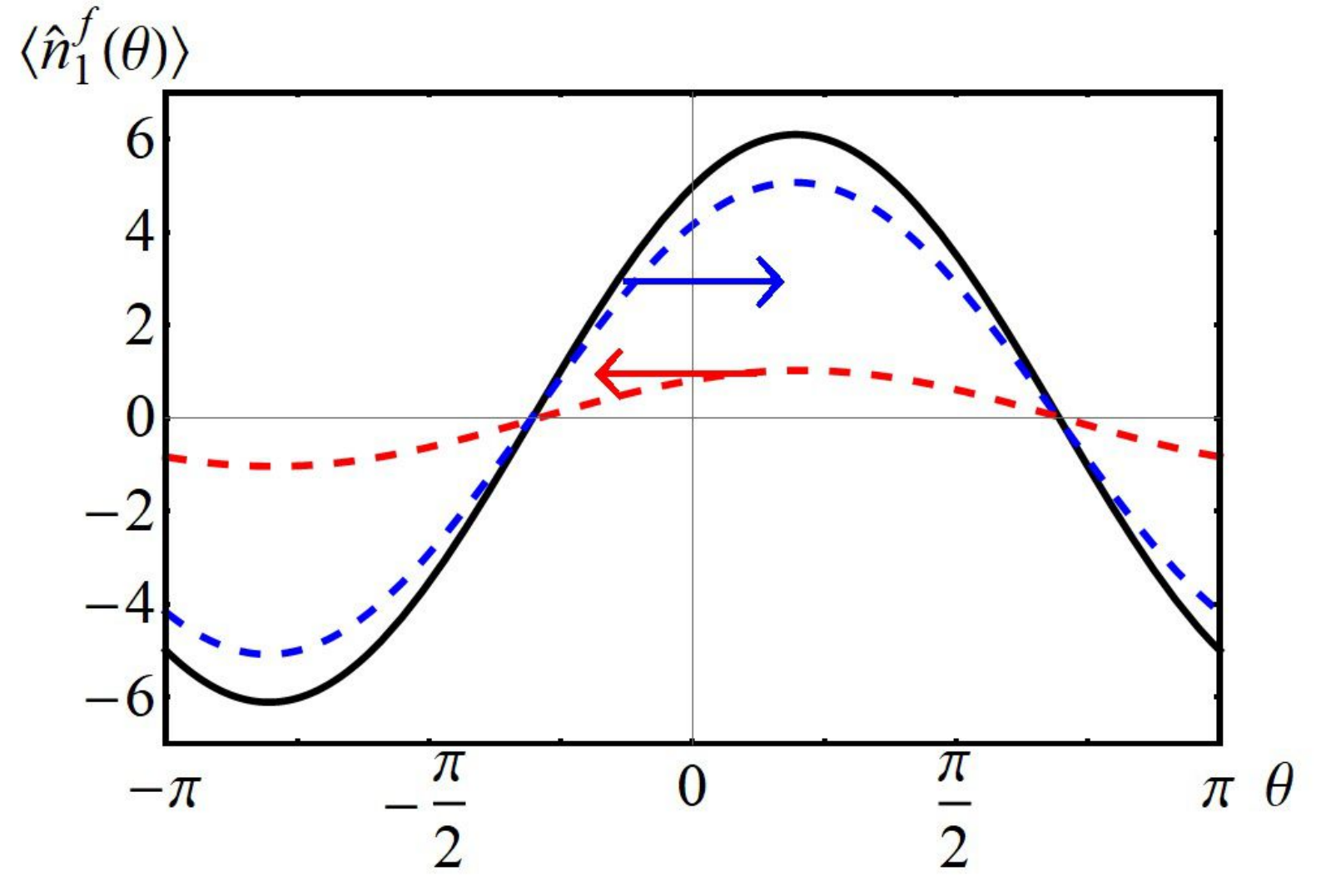}
\label{fig:subfig30}}
\subfloat[Subfigure 2 list of figures text][$\Delta t=8.0\tau$]{
\includegraphics[width=0.24\textwidth]{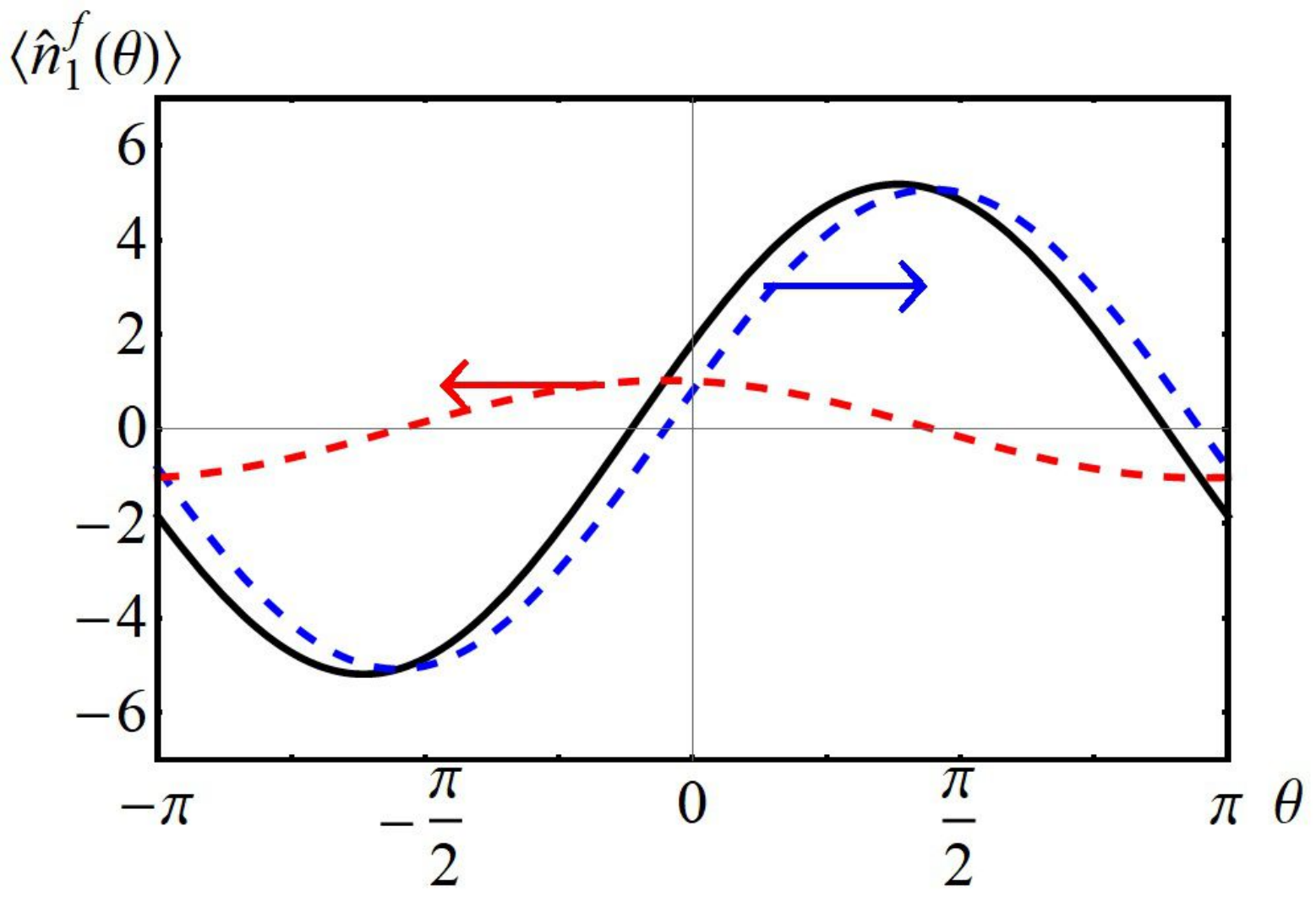}
\label{fig:subfig50}}
\subfloat[Subfigure 2 list of figures text][$\Delta t=11.2\tau$]{
\includegraphics[width=0.24\textwidth]{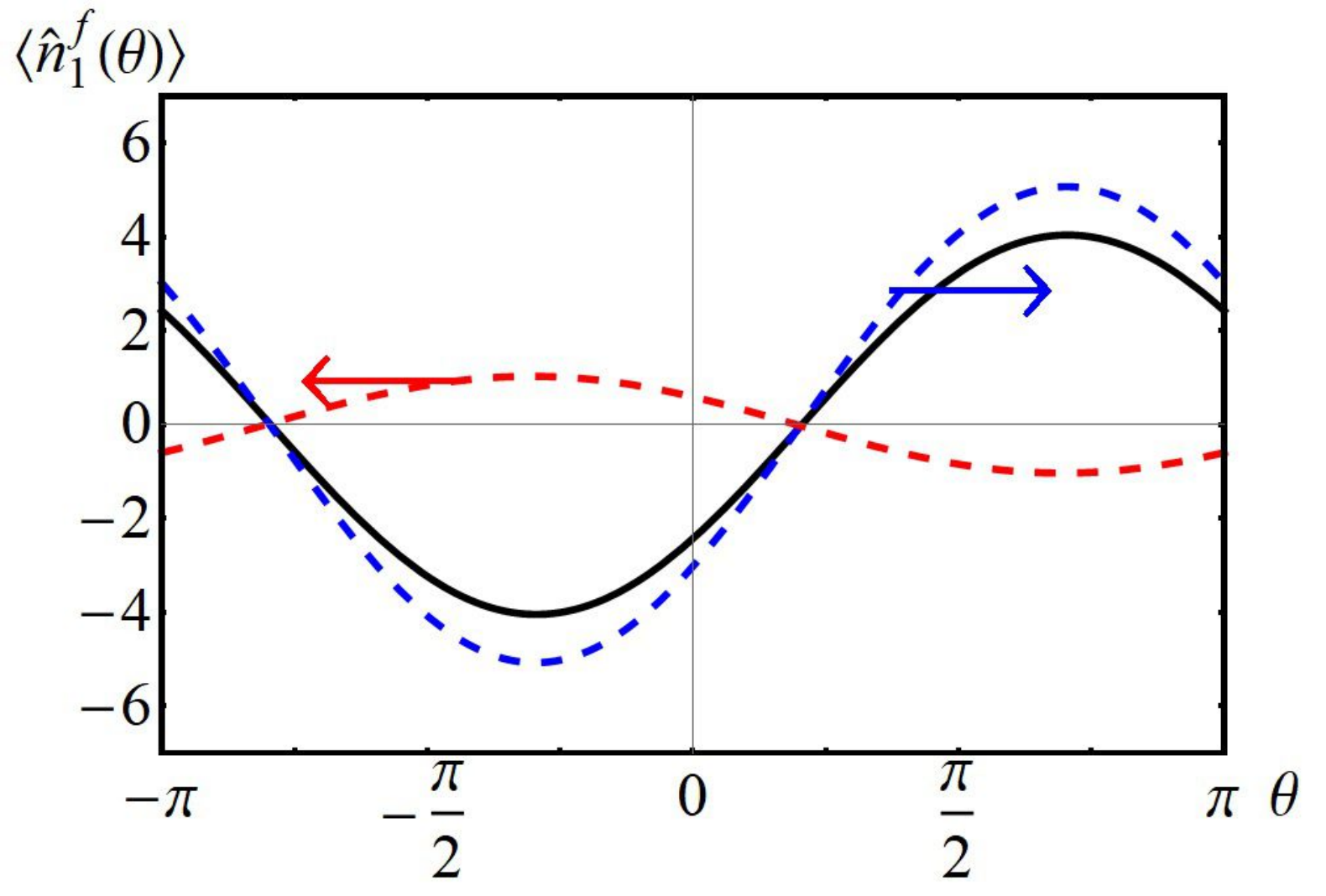}
\label{fig:subfig70}}
\caption{(Color Online) After rapid expansion of the toroidal BEC, an
  initial traveling wave bifurcates into oppositely-oriented traveling
  waves, as depicted in Fig.~\ref{expandingSketch}.  The total density
  of these waves ({\em solid, black}), given by Eq.~(\ref{DensityWaveOut}),
  is a sum of a right-moving wave ({\em dashed, blue, arrow indicating
  direction}), representing the ingoing particles and a left-moving
  wave ({\em dashed, red, arrow indicating direction}) representing the
  dynamically created particles.  For this plot   we studied a
  coherent state characterized by  $\alpha=\sqrt{10}$ and  mode index
  $N=1$, we set the   quantum pressure parameter $\gamma=0.3$, the
  sound speed to be $c=2~$mm/s, the initial radius to be
  $R_{0}=10\mu$m, the duration of expansion to be $t_{f}=10$ms and
  took the timescale governing the trap expansion to be
  $\tau=6.21\text{ms}$.  }
\label{StandingWave}
\end{figure*}

Since we are in the Heisenberg picture, the system wavefunction is always given by Eq.~(\ref{eq:systemwave}),
while the density operators change during the rapid expansion of the ring.   
We start by writing the mode expansion for the initial density operator (before expansion).
To do this, we make use of (\ref{ModeExpn1QP}), and the relation (\ref{RelationDensityPhase}) between
the density $\eta_{n}$ and phase $\chi_{n}$ modes neglecting the quantum pressure corrections
(i.e. $D_{n}\approx 1$ here):
\begin{equation}\label{DensityInitial}
\hat{n}^{i}_{1}(\theta,t) = \curN_{0} \sum_{n=-\infty}^{\infty}\bigg[e^{in\theta}\eta^{\text{in}}_{n}(t)\hat{a}_{n}+e^{-in\theta}\eta_{n}^{\text{in}*}(t)\hat{a}^{\dagger}_{n}\bigg],
\end{equation}
where $\curN_{0}=\sqrt{\frac{U\curV_0}{2\pi\hbar}}$ is the initial
normalization and the \lq in\rq-density modes are
$\eta^{\text{in}}_{n}(t)=i\frac{\hbar}{U}\sqrt{\frac{\omega^{0}_{n}}{2}}e^{-i\omega^{0}_{n}t}$.
Note that we are only neglecting the quantum pressure corrections in the connection between the $\eta_{n}$
and 
$\chi_{n}$ (where they have a small effect) but keeping then in the Mukhanov-Sasaki equaton for
the mode functions (where including the quantum pressure is qualitatively important, as
we have discussed).

Next, we write down the final density operator after expansion, which
has a similar form, but with the \lq out\rq-mode density operators discussed above
(see Eqs.~(\ref{BogoliuobovTransformation}) and (\ref{InverseBogoliubov})):
\begin{equation}\label{DensityFinal}
\hat{n}^{f}_{1}(\theta,t) = \curN_{f}e^{-\frac{t_{f}}{2\tau}\gamma} \sum_{n=-\infty}^{\infty}\bigg[e^{in\theta}\eta^{\text{out}}_{n}(t)\hat{b}_{n}+e^{-in\theta}\eta_{n}^{\text{out}*}(t)\hat{b}^{\dagger}_{n}\bigg],
\end{equation}
with the final normalization $\curN_{f}=\sqrt{\frac{U\curV_f}{2\pi\hbar}}$ and the `out'-density modes
are defined as $\eta^{\text{out}}_{n}(t) = i\frac{\hbar}{U}\sqrt{\frac{\omega^{f}_{n}}{2}}e^{-i\omega^{f}_{n}(t-t_{f})}$. Now we define the average density in the coherent state as:
\begin{equation}
\langle \hat{n}_{1}(\theta,t)\rangle = \langle \alpha, N | \hat{n}_{1}(\theta,t)|\alpha, N \rangle.
\end{equation}
If we take $\alpha\in\mathcal{R}$, then the initial average density can be written as a wave that travels in the 
counterclockwise direction ($+\hat{\theta}$):
\begin{eqnarray}\label{DensityWaveIn}
\langle \hat{n}^{i}_{1}(\theta,t)\rangle & = & -2\sqrt{|N|}\alpha\cdot \sin\bigg[N\theta-\frac{|N|c}{R_{0}}t\bigg],
\end{eqnarray}
where we have set the normalization
$\sqrt{\frac{\hbar c\mathcal{V}_{0}}{4\pi UR_{0}}}$
to unity for simplicity.  

The initial atom density (at $t=0$)  according to
Eq.~(\ref{DensityWaveIn}) is shown in Fig.~\ref{density}  with
a dotted line, describing a counterclockwise traveling density wave (in the
$+\hat{\theta}$ direction).  However, upon evaluating the density expectation value
after expansion, we find that the final average density can be expressed as a
sum of two waves (as illustrated schematically above in Fig.~\ref{expandingSketch}) reflecting phonon creation:   
A counterclockwise wave traveling wave
 (direction $+\hat{\theta}$)  with amplitude
$|\alpha_{n}|$, representing a reduced initial wave, and a smaller
clockwise traveling wave (direction $-\hat{\theta}$)
with an amplitude $|\beta_{n}|$, thus representing the density wave
due to newly created phonons. 
The final time-dependent density is: 
\begin{eqnarray}\label{DensityWaveOut}
 &&\hspace{-.5cm} \langle \hat{n}^{f}_{1}(\theta,t)\rangle  =  -2e^{-\frac{t_{f}}{2\tau}\gamma}|N|^{\frac{1}{2}}\alpha\Bigg(
  |\alpha_{N}|\sin\bigg[N\theta-\frac{|N|c}{R_{f}}\Delta t+\varphi_{\alpha}\bigg] \nonumber \\
&& -  |\beta_{N}|\sin\bigg[N\theta+\frac{|N|c}{R_{f}}\Delta t+\varphi_{\beta}\bigg]  \Bigg),
\end{eqnarray}
where we have again set the normalization $\sqrt{\frac{\hbar c\mathcal{V}_{f}}{4\pi UR_{f}}}$ to unity.
Here, $\Delta t=(t-t_{f})$ is the time elapsed after the expansion has
ended, and $\varphi_{\alpha}=\text{Arg}(\alpha_{N})$ and
$\varphi_{\beta}=\text{Arg}(\beta_{N})$ are, respectively, the phase
associated with the incoming and created particles. The final density
wave is shown in Fig.~\ref{density} at $t=t_{f}$, for various values
of quantum pressure.  In Fig.~\ref{StandingWave} we show the density vs.
angle for increasing values of the elapsed time $\Delta t=(t-t_{f})$.
The total density is shown as a black solid curve, and the contributions due
to the two terms in Eq.~(\ref{DensityWaveOut}), i.e., the abovementioned
counterclockwise and  clockwise contributions, are depicted as blue (rightmoving arrow) and
red (leftmoving arrow) dashed curves, respectively.  Although these two
contributions are not separately measurable, they can be inferred from the time
dependence of the density vs. angle, showing a concrete experimentally testable
signature of particle production in an initial coherent state.

\section{Density Correlations}
\label{sec:DensityCorrelations}
As we have discussed, a rapidly expanding toroidal BEC undergoes a modification of its vacuum, leading
to particle production with amplitude $\beta_n$.  In this section, we show how this is revealed
in  correlations of the density fluctuations (i.e., noise correlations), an experimental probe that has
already been used to study horizons in the context of Hawking radiation~\textcolor{red}{\cite{Steinhauer:2015saa}}.
We start by computing density correlations at zero temperature $T=0$ before generalizing to nonzero temperature.

\subsection{$T=0$ limit}
We assume our initial system, before expansion, is a vacuum of $\hat{a}_{n}$ particles $|0\rangle$. The
appropriate equal-time 
fluctuation correlation function is:
\begin{equation}
  \label{eq:fluctcorr}
  \mathcal{C}(\theta,\theta') = \langle\hat{n}_{1}(\theta,t)\hat{n}_{1}(\theta',t)\rangle -
  \langle\hat{n}_{1}(\theta,t)\rangle\langle\hat{n}_{1}(\theta',t)\rangle,
\end{equation}
where the averages are being taken with respect to the initial vacuum state  $|0\rangle$.  
Then, 
for the initial static BEC before expansion, we make use of
$\langle a_n^\dagger a_{n'}\rangle = 0$ along with 
(\ref{DensityInitial}) and
(\ref{DensityWaveIn}) to calculate the initial noise correlations, obtaining: 
\be
\mathcal{C}^{i}(\theta-\theta') = \curN_0^2 \sum_{n=-\infty}^\infty
|\eta^{\text{in}}_{n}(t)|^2
{\rm e}^{in(\theta-\theta')}.
\ee
Since the mode functions in the summand $|\eta^{\text{in}}_{n}(t)|^2 = \frac{\hbar^{2}}{2U^{2}}\omega_n^0 \propto |n|$,
this sum is divergent and must be regularized.  To implement this regularization, we note that this divergence
comes from the fact that we have taken a
long-wavelength (low energy) approximation.  The linear-in-$n$ energy dependence of these modes must become
quadratic, as in the conventional Bogoliubov approximation,  for sufficiently large
$|n|\geq n_c \equiv \frac{2cM R_0}{\hbar}$.  We account for this this by replacing the linear summand with
the result following from Bogoliubov theory, which gives:
\bea
\nonumber
\mathcal{C}^{i}(\theta-\theta') &=& \curN_0^2 \frac{\hbar^2 c}{2U^2 R_0} n_c\sum_{n=-\infty}^\infty \frac{n^2}{\sqrt{n^2(n^2+n_c^2)} }
        {\rm e}^{in(\theta-\theta')},
        \\
         &=&  \frac{1}{2\pi} n_0 \curV_0 \sum_{n=-\infty}^\infty  \frac{n^2}{\sqrt{n^2(n^2+n_c^2)} }
        {\rm e}^{in(\theta-\theta')},
\eea
where in the second line we inserted our formulas for $\curN_0$, $n_c$, and $c$ (at the initial radius $R_0$)
to simplify the prefactor.  This result is precisely what one would obtain for a 1D BEC, at $T=0$, within standard Bogoliubov theory~\cite{Imambekov}.
The final sum is convergent, although it has a
delta-function piece that we can isolate with the Poisson summation formula to arrive at 
\bea
\label{NoiseInitial}
\mathcal{C}^{i}(\theta-\theta')  &=&
n_0 \curV_0 \Big( \delta(\theta-\theta') + \curS(\theta-\theta')\Big) ,
\\
\curS(\theta) &\equiv & \frac{1}{2\pi}\sum_{n=-\infty}^\infty \Big[\frac{n^2}{\sqrt{n^2(n^2+n_c^2)}}  -1\Big]
        {\rm e}^{in\theta},~~~
\eea
for the noise correlations before expansion.

The result for the  final regularized noise correlations after expansion follow similarly, and
can be written in the following manner:
\begin{eqnarray}
  \nonumber 
  &&\mathcal{C}^{f}(\theta-\theta') =
  n_{0}\mathcal{V}_{f}e^{-\frac{t_{f}}{\tau}\gamma}\Big(\delta(\theta-\theta')
\\
&& \qquad \qquad +\mathcal{S}(\theta-\theta') +\mathcal{C}^{\text{sub}}(\theta-\theta')\Big),
  \label{NoiseFinal}
\end{eqnarray}
where we defined the function
\bea
&&\mathcal{C}^{\text{sub}}(\theta)  \equiv   \frac{1}{2\pi}\sum_{n=-\infty}^{\infty}\frac{n^{2}}{\sqrt{n^{2}(n^{2}+n_{c}^{2})}} \nonumber \\
&&\qquad  \times  2\Big[|\beta_{n}|^{2}-\text{Re}\Big(\alpha_{n}\beta^{*}_{n}e^{-2i\omega^{f}_{n}\Delta t}\Big)\Big] e^{in\theta},
\eea
which, as can be seen by comparing to Eq.~(\ref{NoiseInitial}), is an additional contribution after the rapid expansion of
the ring BEC.  Here, the superscript \lq\lq sub\rq\rq\ indicates
that this is the subtracted noise correlator, i.e., the difference of the final and initial
normalized correlators.
Note that this contribution depends on $\Delta t=(t-t_{f})$, the time elapsed after
expansion, so that the summand exhibits oscillatory behavior as a function of $\Delta t$.
However, we find that the summand is well approximated by time-averaging over one period ($2\pi/\omega_n$) of these oscillations,
which eliminates the interference term $\alpha_{n}\beta^{*}_{n}$ and 
yields
\begin{equation}
  \label{SubtractedCorrelator}
  \mathcal{C}^{\text{sub}}(\theta) \simeq \frac{1}{\pi}\sum_{n=-\infty}^{\infty}\frac{n^{2}}{\sqrt{n^{2}(n^{2}+n_{c}^{2})}}
  |\beta_{n}|^{2} e^{in\theta},
\end{equation}
 for the subtracted noise correlations.

\begin{figure}[h]
\centering
\includegraphics[width=0.5\textwidth]{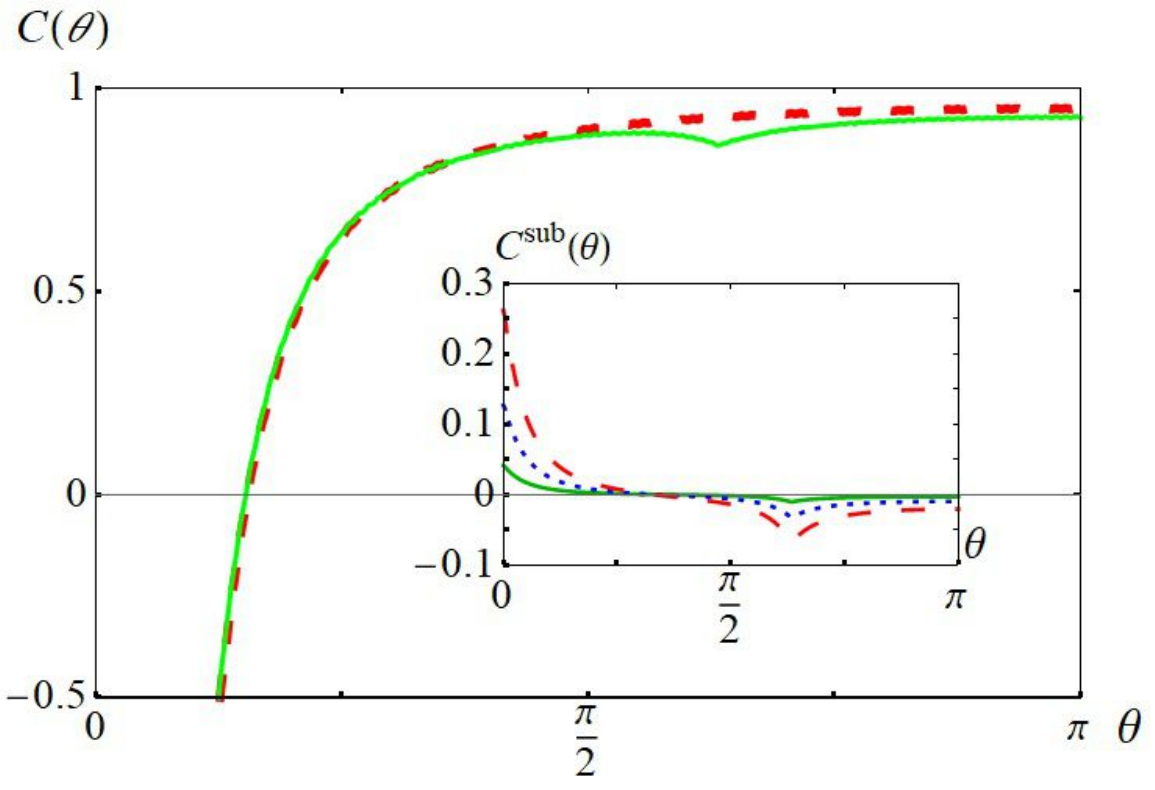}
\caption{(Color Online) Main: Noise correlations $\mathcal{C}(\theta)$
  at zero temperature, before  ({\em dashed, red\/},  given in Eq.~(\ref{NoiseInitial}), and
  after ({\em solid, green\/}, given in Eq.~(\ref{NoiseFinal})) expansion, with
  quantum pressure
  $\gamma=0.5$.  In each, we focused on $\theta\neq 0$ (dropping the
  delta-function piece) and dropped overall prefactors.
  The correlations after expansion show a cusp like
  feature that we regard as a signature of particle creation.  Inset:
  To emphasize the cusp, we plot the difference between these curves,
  the subtracted noise correlations 
  $\mathcal{C}^{\text{sub}}(\theta-\theta')$
  plotted with respect to the angle difference $\theta$.  Here we chose
  three values of quantum pressure parameter 
  $\gamma=0.2$ ({\em solid, green}) and $\gamma=0.35$ ({\em
    short-dashed, blue}) and $\gamma=0.5$ ({\em long-dashed, red}).
  For both plots, we took the speed of sound to be $c=2~$mm/s, the
  initial radius to be $R_{0}=10\mu$m, the duration of expansion to be
  $t_{f}=10$ms, the timescale governing the trap expansion to be
  $\tau=6.21\text{ms}$, and  $n_{c}=10$.
}
\label{Parts}
\end{figure}

The expressions (\ref{NoiseInitial}) and (\ref{NoiseFinal}) for the
initial and final  noise correlations  have a dirac-delta function
that is divergent at $\theta=\theta'$.  In plotting these functions,
we drop this piece, and set the prefactors (i.e.,
$n_{0}\mathcal{V}_{0}/(2\pi)$ and $n_0 \curV_{f}{\rm
  e}^{-\frac{t_{f}}{\tau}\gamma}/(2\pi)$) to unity to simplify comparing
 the noise before and after expansion.  The main part of
Fig.~\ref{Parts} shows this comparison, with the initial case being a
red dashed line and the final case being a solid green line.  We see
that each  case is dominated by a large (though finite) negative contribution at equal
angles $(\theta\to 0)$.  We regard such  anti-correlations as
reflecting the  repulsion of bosonic atoms at short
distances~\textcolor{red}{\cite{Carusotto:2008ep,Glauber}}. For
larger angular separations, the correlations gradually flatten
out~\textcolor{red}{\cite{Schellekens}}, except for the appearance of
a cusp feature at nonzero angle in the final noise correlations.
This cusp clearly represents a signature of the phonon
creation, proportional to $|\beta_n|^2$, that we have discussed above.
In the inset,
we plot the subtracted part Eq.~(\ref{SubtractedCorrelator}) for
three values  of the quantum pressure parameter: $\gamma=0.2$ (solid
green), $\gamma=0.35$ (short-dashed blue) and $\gamma=0.5$
(long-dashed red). This shows that the magnitude of the cusp increases
with increasing quantum pressure, although the cusp location is
independent of $\gamma$. We do find that the cusp location as a function of angle
increases with increasing expansion time $t_f$,  asymptotically approaching $\theta =\pi$ for
large $t_f$.   
%

To better understand the origin of the cusp  in Fig.~\ref{Parts}, and connect it to particle
production during expansion of the ring BEC,
we differentiate (\ref{SubtractedCorrelator}) to get:
\begin{equation}
  \label{differentiateC}
\frac{d}{d\theta}\mathcal{C}^{\text{sub}}(\theta) =
-\frac{2}{\pi}\sum_{n=1}^{\infty}\frac{n^{2}|\beta_{n}|^{2}}{\sqrt{n^{2}+n_{c}^{2}}}\sin(n\theta).
\end{equation}
A good approximation to this sum results if we Taylor expand the creation parameter $\beta_{n}$ for small damping $\gamma\ll
1$ and large mode index $n\gg 1$, which gives 
$n^{2}|\beta_{n}|^{2}\approx\big(\frac{\gamma}{4}\big)^{2}\big(\frac{c\tau}{R_{0}}\big)^{-2}\Big[1+a^{2}-2a\cos\big(2n\theta_{\rm
    H}\big)\Big]$, where $a=e^{\frac{t_{f}}{\tau}}$ is the scale
factor and $\theta_{\rm H}=\frac{c\tau}{R_{0}}\big(1-a^{-1}\big)$ is
the angular horizon size at the end of expansion. Upon plugging this
into (\ref{differentiateC}), we end up with three sums that we write as:
\begin{eqnarray}
&&\frac{1}{f(\gamma)}\frac{d}{d\theta}\mathcal{C}^{\text{sub}}(\theta) = 
a\sum_{n=1}^{\infty}\frac{\sin(n(\theta-2\theta_{\rm H}))}{\sqrt{n^{2}+n_{c}^{2}}} \nonumber \\
&& +~a\sum_{n=1}^{\infty}\frac{\sin(n(\theta+2\theta_{\rm H}))}{\sqrt{n^{2}+n_{c}^{2}}} 
 -  (1+a^{2})\sum_{n=1}^{\infty}\frac{\sin(n\theta)}{\sqrt{n^{2}+n_{c}^{2}}},~~~~~~~\label{differentiateC1}
\end{eqnarray}
where
$f(\gamma)=\frac{\gamma^{2}}{8\pi}\big(\frac{c\tau}{R_{0}}\big)^{-2}$.  We now analyze the angle
dependence of this quantity.  For $\theta>0$, the
first sum is the dominant one, and approximating it by an integral
we get
\bea
&&\hspace{-.5cm}\sum_{n=1}^{\infty}\frac{\sin(n(\theta-2\theta_{\rm H}))}{\sqrt{n^{2}+n_{c}^{2}}}\simeq
\int_{0}^{\infty}dx\frac{\sin\big(n_{c}(\theta-2\theta_{\rm H})x\big)}{\sqrt{1+x^{2}}}, \nonumber
\\
&&\hspace{-.5cm}\simeq 
\frac{\pi}{2}\big[\Theta(\theta-2\theta_{\rm
    H})I_{0}\big(n_{c}(\theta-2\theta_{\rm H})\big)-L_{0}\big(n_{c}(\theta-2\theta_{\rm
    H})\big)\big],~~~~~~
\eea
where the integration variable $x=n/n_{c}$, $I_{0}(x)$ is the modified Bessel function of the
first kind, $L_{0}(x)$ is the modified Struve function and $\Theta(x)$
is the unit step function. The appearance of the latter means that
the derivative is discontinuous at $\theta=2\theta_{\rm H}$, implying that this determines the
angular position of the cusp. We therefore conclude that the cusp in
the correlation function is determined by the angular horizon size:
\begin{equation}\label{LocationCusp}
\theta_{\text{cusp}} = 2\theta_{\text{H}} = 2\frac{c\tau}{R_{0}}\Big(1-e^{-\frac{t_{f}}{\tau}}\Big).
\end{equation}
This result shows that the cusp location is independent of the damping
parameter $\gamma$, as we saw in Fig.~\ref{Parts}. It also explains why the cusp moves away from the
origin, eventually slowing down as it approaches $\theta=\pi$, as the
duration of expansion $t_{f}$ increases. The negative correlation at
the cusp signifies creation of phonons that anti-bunch as they are
created in pairs that move away from each other with opposite
momenta~\textcolor{red}{\cite{Birrell:1982ix,Mukhanov:2007zz,Parker and  Toms,Fulling:1989nb}}. This is similar to the `tongue'-like features that
were numerically observed in an acoustic black
hole~\textcolor{red}{\cite{Carusotto:2008ep}}. \\ 

\begin{figure}[h]
\centering
\includegraphics[width=0.5\textwidth]{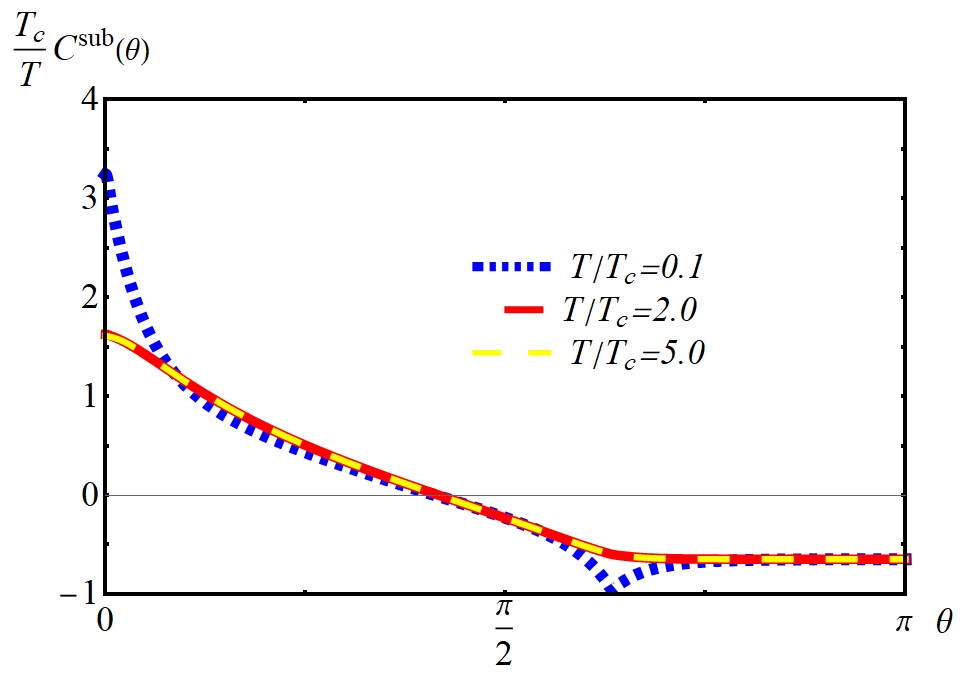}
\caption{ (Color Online) Subtracted noise correlations
  $\frac{T_{c}}{T}\mathcal{C}^{\text{sub}}(\theta)$ plotted with
  respect to the angle difference $\theta$, at finite temperatures
  given in Eq.~(\ref{SubtractedCorrelatorT}). In each, we focused on
  $\theta\neq 0$ (dropping the delta-function piece) and dropped
  overall prefactors.
  For
  low temperature $T/T_{c}=0.1$ ({\em short-dashed, blue}), the
  correlations after expansion show a cusp like feature as in
  Fig.~\ref{Parts}. For high temperatures $T/T_{c}=2$ ({\em solid,
    red}) and $T/T_{c}=5$ ({\em long-dashed, yellow}), we get similar
  correlations with a kink at the same location where the
  cusp appears. The appearance of the kink and the scaling behavior
  with temperature is a signature of particle creation.
}
\label{CsubFiniteTemp}
\end{figure}

\subsection{$T\neq 0$ regime}
A key question concerns the fate of the cusp feature at finite temperatures.  To explore this, 
we extend the preceding analysis to finite temperatures by assuming the initial unexpanded ring is characterized by a Bose
distribution of quasiparticles. We therefore evaluate the averages in Eq.~(\ref{eq:fluctcorr}) using: 
\be
  \label{ThermalAverages}
\langle\hat{a}_{n}^{\dagger}\hat{a}_{n'}\rangle  =  \delta_{n,n'}n_{B}(E_{n})
\ee
where $n_B(x) =(e^{\beta x}-1)^{-1}$ is the Bose-Einstein distribution with the chemical potential set to zero
  for massless phonons, $\beta^{-1}=k_{B}T$ and $E_{n}=\sqrt{\epsilon_{n}(\epsilon_{n}+2Un_{0})}$ is the Bogoliubov dispersion.
Following similar steps as in the preceding, we find for the subtracted correlation function: 
  \begin{equation}\label{SubtractedCorrelatorT}
\mathcal{C}^{\text{sub}}(\theta) \simeq \frac{1}{\pi}\sum_{n=-\infty}^{\infty}\frac{n^{2}}{\sqrt{n^{2}(n^{2}+n_{c}^{2})}}\big(1+2n_{B}(n,T)\big)|\beta_{n}|^{2} e^{in\theta},
\end{equation}
  which is similar to $(\textcolor{red}{\ref{SubtractedCorrelator}})$ except for the appearance of a
  Bose-Einstein distribution function $n_{B}(n,T)$ defined as
  \begin{eqnarray}
n_{B}(n,T) & = & \Big(e^{f(n)\frac{T_{c}}{T}}-1\Big)^{-1}, \nonumber \\
f(n) & \equiv & \Big|\frac{n}{n_{c}}\Big|\bigg(1+\frac{n^{2}}{n^{2}_{c}}\bigg)^{1/2}
\end{eqnarray}
  where the temperature scale is defined as $T_{c} \equiv \frac{n^{2}_{c}\hbar^{2}}{2MR^{2}k_{B}}$.
  
 In Fig.~\ref{CsubFiniteTemp}, we plot the subtracted noise correlator 
 (\textcolor{red}{\ref{SubtractedCorrelatorT}}), divided by $T/T_c$,
 for three different
 temperatures $T/T_{c}=0.1$, $T/T_{c}=2$ and $T/T_{c}=5$.
For these curves we chose the quantum pressure to be $\gamma=0.5$, the
  speed of sound to be $c=2~$mm/s, the radius to be $R=10\mu$m, the
  duration of expansion to be $t_{f}=10$ms, the timescale governing
  the expansion of the trap to be $\tau=6.21\text{ms}$ and
  $n_{c}=10$. Using these parameters we get $T_{c}=10~$nK for the relevant
  temperature scale.

These curves show that, 
  while the cusp-like feature is still present at low $T$, it is smoothed out at higher $T$.
  The higher $T$ curves instead exhibit a change of slope (i.e., a kink) at the horizon location.
  In fact, the two higher temperature curves in Fig.~\ref{CsubFiniteTemp} overlap. This is due to the fact
  that they  are well approximated by taking the high-temperature limit of the Bose function, replacing
  $n_B(n,T) \approx T/(fT_c)$ in Eq.~(\ref{SubtractedCorrelatorT}) to obtain for the  subtracted correlator:
\begin{eqnarray}\label{CsubApprox}
\mathcal{C}^{\text{sub}}(\theta) & \approx & \frac{2T}{\pi T_{c}}\sum_{n=-\infty}^{\infty}\frac{|\beta_{n}|^{2}}{1+(n/n_{c})^{2}}e^{in\theta}.
\end{eqnarray}
The approximate result Eq.~(\ref{CsubApprox}) shows that, with increasing temperature, the angle-dependent subtracted noise
correlations scale linearly with $T$, but with an angle-dependence that still reflects the horizon location.

From these results, we can outline two experimental procedures to identify stimulated
particle creation in an experiment measuring density correlations.  Firstly, an experiment
that measures correlations at different temperatures (but for the same expansion time) would
find an approximate collapse of the curves, when plotted as normalized in Fig~\ref{CsubFiniteTemp}.
This probes the predicted linear scaling with temperature.  
Secondly,  experiments at fixed temperature but varying expansion time would detect
a change in the location of the angle-dependent kink, representing the moving horizon.  
These results demonstrate that correlations in the
  density fluctuations show a clear signature of particle production
  in a rapidly expanding toroidal BEC, showing another way that cold atom experiments can probe
  inflationary physics.

\section{Concluding Remarks}
\label{sec:concl}
In this paper, we have explored how an exponentially expanding thin
toroidal Bose-Einstein condensate can reproduce the various features
of primordial cosmological inflation.
Our work was inspired by recent experimental and theoretical work by Eckel and collaborators
who studied inflationary physics in a ring-shaped BEC~\textcolor{red}{\cite{Eckel:2017uqx}}.
These authors observed experimentally (and confirmed theoretically) the redshifting of phonons due to the rapid expansion of
this analogue 1D universe, a damping of
phonon modes due to Hubble friction, and evidence of the preheating phenomena predicted to
occur at the end of inflation.

A central finding of our work is that quantum pressure effects, even if they
are quantitatively small, can have important implications for the dynamics of
expanding toroidal BEC's.  Such quantum pressure effects modify the
Mukhanov-Sasaki equations for phonon modes in a fundamental way,
with the resulting solutions exhibiting damping and redshift, 
just like in inflationary cosmology.   We found that this damping 
is responsible for the change of the vacuum state of the
fluctuations, which ultimately leads to the dynamical generation of
phonons. This is the analogue of particle production in the early
universe. As a result, if the perturbations start in a coherent state,
the ring expansion forces them to bifurcate into two density waves
that propagate opposite to each other, leading to a complex time-dependent
density wave in the toroid. 
This phonon generation also manifests itself in the density-density
noise correlations as a cusp-like feature that tracks the horizon
size.  Both of these results are clear signatures of particle creation
and can be verified experimentally. 
  However, it is
  important to note that, within the gravitational analogy viewpoint,
  the model we consider here is a very special one i.e. a
  quasi-one-dimensional toroidal BEC, the small width of which led us
  to consider the short-distance corrections due to quantum
  pressure. More general models could be realized experimentally that
  can give different results and interpretations
  \textcolor{red}{\cite{Wittemer:2019kds}}.

We make note of two things. Firstly, a simple harmonic oscillator with
a time dependent frequency~\textcolor{red}{\cite{Mukhanov:2007zz}},
can also exhibit a change of vacuum
states leading to particle creation. Secondly, the fact that no
particle creation happens for $\gamma=0$ is related to the presence of an
adiabatic invariant in 1D~\textcolor{red}{\cite{Eckel:2017uqx}}.
Unlike other settings where quantum pressure effects are unimportant for
particle production (for example, the  Sakharov oscillations measurements of Hung et
al~\textcolor{red}{\cite{Hung:2012nc}}), 
in the present case of a one-dimensional
BEC, where we have assumed a time independent speed of sound $c$,
it is essential to have a nonzero $\gamma$ for the damping of
modes.
We emphasize that taking  $c$ to be constant is a simplifying approximation, and that
a rapid time-dependent variation of $c$ would also lead to phonon production.
Within the preceding approximations, our work shows that
quantum pressure is  essential to achieve particle production, as seen from the
form of Eq.~(\ref{MSeasy}) and its solution Eq.~(\ref{Eq:chiproper}).

   Such damping is also crucial within inflationary theory, where
   the horizon size is approximately fixed.
   During this period, some modes  (which are due to
   spontaneously created particles) get stretched out of the
   horizon and thus freeze. However, other modes never exit the horizon and thus
   undergo damping. When inflation ends, the horizon again
   expands and starts enveloping these frozen modes, that re-enter the
   horizon and distribute the available matter and radiation. This way,
   inflation provides a mechanism through which vacuum fluctuations in
   the early universe manifest themselves later in the form of
   distribution of galaxies and the CMB anisotropies. In a BEC, the
   quantum pressure is essential for damping and particle creation. As
   a result, here too the same mechanism of horizon exit and freezing
   of modes is happening. In this sense, the quantum pressure terms provide us with
   an analogue of the inflationary mechanism in a BEC.

Future studies could look into other types of expansion rates like the
ones arising from the quadratic or the Starobinsky models of
inflation. A further possibility as mentioned
in~\textcolor{red}{\cite{Eckel:2017uqx}}, could be to study how
causally disconnected regions recombine. This could potentially help
cosmology experiments to observe physics beyond our current horizon.
  Another possibility as discussed in \cite{Cha:2016esj}, is to see whether cold atomic systems can be used to
  study the trans-Planckian era that happened before inflation, which is believed to entail quantum effects of gravity.

\section{Acknowledgements}
The authors are grateful to 
Ivan Agullo for useful comments.~AB acknowledges financial support from the Department of Physics and Astronomy at LSU.
DV acknowledges financial support from LSU.
DES acknowledges financial support from National Science Foundation Grant No.
DMR-1151717.

\appendix
\section{Dynamics of an expanding toroidal BEC}
\label{appendix}
In this section we study BEC's in the presence of an expanding
toroidal-shaped trap given by Eq.~(\ref{eq:veeofr}).  Our aim is to
understand the background solution on which phonon excitations
propagate.  
For this task we study the time-dependent Gross-Pitaevskii equation (GPE)
\bea
&&i\hbar \frac{\partial}{\partial t} \Phi_0(\br,t) =  - \frac{\hbar^2}{2M} \nabla^2 \Phi_0(\br,t) +
\big(V(\br,t) - \mu\big) \Phi_0(\br,t)
\nonumber \\
&&\qquad 
+ U|\Phi_0(\br)|^2 \Phi_0(\br,t).
\label{eq:TDGPE}
\eea
Writing $\Phi_0(\br,t) = \sqrt{n_0(\br,t)}{\rm e}^{i\phi_0(\br,t)}$, with $n_0$ the density and $\phi_0$ the superfluid
phase, we obtain:
\bea
\label{Eq:gpeAN1}
&&- \hbar \partial_t\phi_0(\br,t) = - \frac{\hbar^2}{2M \sqrt{n_0(\br,t)}} \nabla^2 \sqrt{n_0(\br,t)}
 \\
&&\qquad
+
\frac{\hbar^2}{2M } (\grad\phi_0(\br,t) )^2
+ V(\br,t) -\mu+ Un_0(\br,t) ,\nonumber
\\
&&
\partial_t n_0(\br,t) = - \frac{\hbar}{M}
\grad\cdot \big(n_0(\br,t)\grad \phi_0(\br,t)\big),
\label{Eq:gpeAN2}
\eea
A key question is whether the superfluid velocity,
$\bv(\br,t) = \frac{\hbar}{M}\grad\phi_0(\br,t)$, is equal to the
radial ring velocity $\hat{\rho}\dot{R}(t)$.  Before analyzing this, we recall the simpler case of
a {\em homogeneously translated\/} trap moving at constant velocity $\bv_T$.
In this case, which can be described by a  trapping potential $V(\br,t) = V(\br-\bv_T t)$,
Galilean invariance~\textcolor{red}{\cite{Yukalov}}
ensures that a solution to Eqs.~(\ref{Eq:gpeAN1}) and Eqs.~(\ref{Eq:gpeAN2}) always
exists with superfluid velocity $\bv = \bv_T$ and density $n(\br)$ static in the moving frame.  That is, we can always
boost to a moving reference frame in which the single-particle potential is static.  

In the case of present interest, however, a toroidal expanding ring described by the
trapping potential Eq.~(\ref{eq:veeofr}), the lack of Galilean invariance
means that we cannot find such a simple exact solution with $\bv= \hat{\rho}\dot{R}(t)$.
In the following, we investigate whether such a relation holds approximately
under the conditions of the experiment.  To do this, we
take the gradient of both sides of Eq.~(\ref{Eq:gpeAN1}), and use the definition of
the superfluid velocity, to obtain the Euler equation:
\bea
  \nonumber 
&&- M \partial_t \bv =  \grad\Big[ - \frac{\hbar^2}{2M \sqrt{n_0(\br,t)}}
  \nabla^2 \sqrt{n_0(\br,t)} 
  \\
  &&\qquad 
    + \frac{1}{2}M\bv^2
  + V(\br,t) -\mu+ Un_0(\br,t) \Big].
\label{eq:euler}
\eea
We now invoke the Thomas-Fermi (TF) approximation~\textcolor{red}{\cite{BaymPethick}} by neglecting the Laplacian term
in square brackets on the right of Eq.~(\ref{eq:euler}).  Then, we plug our
assumed solution $\bv= \hat{\rho}\dot{R}(t)$ into the left side,
which leads to $-M \partial_t \bv  = -M \ddot{R}\rhoh$, allowing us to find the following
result for the TF density of a BEC in an expanding toroid:
\bea
\nonumber
&&n_0(\br) = \frac{1}{U}\Big( \mu(t) - \frac{1}{2}M\omega_z^2 z^2 -
\lambda |\rho -R|^n  - M\ddot{R} (\rho-R)\Big)
\\
&&\times\Theta\Big(\mu(t) - \frac{1}{2}M\omega_z^2 z^2 - \lambda |\rho -R|^n
 - M\ddot{R} (\rho-R)
\Big),
\label{eq:nvsrhoagain}
\eea
obtained by integrating both sides of Eq.~(\ref{eq:euler}) with respect to
$\rho$.  Note we also plugged in $V(\br,t)$ from Eq.~(\ref{eq:veeofr}), and an
overall constant of integration was chosen so that the $\rho$
dependence of Eq.~(\ref{eq:nvsrhoagain}) is via the combination $\rho-R(t)$
(although we suppressed the time argument in $R$ for brevity).  

The chemical potential in Eq.~(\ref{eq:nvsrhoagain}) is determined by satisfying
the fixed number constraint $N = \int d^3 r\, n_0(\br)$, with $N$ the total
boson number.
In this integration, the term proportional to $(\rho-R(t))$ will approximately
vanish, with the other terms in Eq.~(\ref{eq:nvsrhoagain}) determining
the TF radii in the $z$ and $\rho$ directions, which are given by:
\bea
R_z &=& \sqrt{\frac{2\mu(t)}{M\omega_z^2}},
\\
R_\rho &=& \Big( \frac{\mu(t)}{\lambda}\Big)^{1/n}.
\eea 
With these definitions, the density is given by:
\bea
&&n_0\simeq \frac{\mu(t)}{U}
\Big[1-  \frac{z^2}{R_z^2} - \frac{1}{R_\rho^n}|\rho -R|^n  -\frac{M\ddot{R}}{\mu(t)} (\rho-R)\Big]
\nonumber
\\
&&\times\Theta\Big(1-  \frac{z^2}{R_z^2} - \frac{1}{R_\rho^n}|\rho -R|^n   -\frac{M\ddot{R}}{\mu(t)} (\rho-R)   \Big),
\label{eq:aeght}
\eea
describing a peak in the atom density that approximately follows the expanding ring.  
 The large value of the exponent $n$ implies a ``flatness'' to the
density profile in the radial direction, i.e., a weak dependence of the density on $\rho$.
We note that in the Eckel et al experiments the exponent $n\simeq 4$, although we'll keep it
general in this section. 

The system chemical potential $\mu(t)$ is determined by
the requirement of a fixed total particle number $N$ during expansion.  Since the density at the
center of the toroid (i.e. at $\rho = R(t)$ and $z=0$) is proportional to $\mu(t)$ in Eq.~(\ref{eq:aeght}),
and the toroid volume
is proportional to $R_zR_\rho R(t)$, then the fixed number constraint leads to the estimate
\be
N \propto \mu(t) R_z R_\rho R(t)\propto \mu(t)^{\frac{3n+2}{2n}}R(t), 
\ee
which implies the chemical potential satisfies
\be
\label{eq:mures}
\mu(t) \propto  R(t)^{-\frac{2n}{3n+2}},
\ee
with exponent $\gamma \equiv \frac{2n}{3n+2}\simeq \frac{4}{7}$.
Thus, during
expansion, the chemical potential (and central density) decrease with increasing time.
Note that since the sound velocity  $c\propto \sqrt{n_0}$, this result implies that
the sound velocity scales with toroidal radius as $c\propto R(t)^{-\frac{1}{2}\gamma}$, or $R(t)^{-2/7}$ for the case of
$n=4$~\textcolor{red}{\cite{Eckel:2017uqx}}.

We have found that a solution with $\bv \simeq \dot{R}(t)\rhoh$  can approximately satisfy the Euler equation and yields
a time-dependent chemical potential in the number constraint
equation.  The next step is to examine 
the continuity equation, Eq.~(\ref{Eq:gpeAN2}), which is:
\be
\label{Eq:continuitynewd}
\partial_t n_0 = - \grad n_0 \cdot \bv - n_0 \grad\cdot \bv .
\ee
We now analyze Eq.~(\ref{Eq:continuitynewd}) without assuming $\bv \simeq \dot{R}(t)\rhoh$, but only the TF
density profile result Eq.~(\ref{eq:nvsrhoagain}).  
To simplify the left side of Eq.~(\ref{Eq:continuitynewd}),  we note that Eq.~(\ref{eq:nvsrhoagain}) implies that 
the partial time derivative of $n_0$ satisfies:
\bea
&&\partial_t n_0  = - \dot{R}(t) \rhoh \cdot \grad n_0(\rho,z,t)
\\
&&\nonumber 
+ \frac{1}{U} \Big( \partial_t \mu(t)
- M(\rho-R(t)) \dddot{R}(t) \Big).
\eea
Henceforth we drop the final term on the right side, since it is small 
in the regime $\rho \to R(t)$.  Plugging this into the left side of Eq.~(\ref{Eq:continuitynewd}) gives
\be
 - \dot{R}(t) \rhoh \cdot \grad n_0+\frac{1}{U}\partial_t \mu(t)  = - \grad n_0 \cdot \bv - n_0 \grad\cdot \bv .
\ee
We now analyze this equation in the regime of $\rho \simeq R(t)$.  From  Eq.~(\ref{eq:nvsrhoagain}),
we find that the gradient of $n_0$ is a constant at $\rho \to R$ and is given by: 
\be
\rhoh \cdot \grad n_0\Big|_{\rho\to R(t)} = -\frac{M\ddot{R}(t)}{U}.
\ee
Plugging this in to the continuity equation, using our result for $\mu$ (which implies
$\partial_t \mu = - \gamma \frac{\dot{R}}{R}\mu$), and cancelling an overall factor of $1/U$, we find:
\be
\label{aftercancelling}
M \dot{R} \ddot{R} - \mu \gamma \frac{\dot{R}}{R} = M \ddot{R} v - \mu\grad\cdot \bv.
\ee
Now we take account of the fact that our system exhibits a rapid growth of $R$ with increasing $t$.
During this expansion, $\mu$ decreases slowly according to Eq.~(\ref{eq:mures}), while $\frac{\dot{R}}{R}$ is
$\curO(1)$ (e.g. for exponential growth).  This  implies that the first terms on the right and left
sides of Eq.~(\ref{aftercancelling}) are much larger than the second terms on 
the left and right sides.  Dropping the subleading terms, we finally get:
\be
M \dot{R} \ddot{R} = M \ddot{R} v,
\ee
or $v = \dot{R}$, consistent with our original assumption. This shows that, within the preceding
approximations, a rapidly expanding toroidal BEC indeed exhibits a radial superfluid velocity
$\bv = \dot{R}\rhoh$.

\bibliography{Bibliography.bib}

\end{document}